\makeatletter
\pdfoutput=1
\makeatother
\documentclass[12pt]{article}\usepackage[]{graphicx}\usepackage[]{xcolor}
\makeatletter
\def\maxwidth{ %
  \ifdim\Gin@nat@width>\linewidth
    \linewidth
  \else
    \Gin@nat@width
  \fi
}
\makeatother

\definecolor{fgcolor}{rgb}{0.345, 0.345, 0.345}

\usepackage{framed}
\makeatletter
 {\par\unskip\endMakeFramed%
 \at@end@of@kframe}
\makeatother

\definecolor{shadecolor}{rgb}{.97, .97, .97}
\definecolor{messagecolor}{rgb}{0, 0, 0}
\definecolor{warningcolor}{rgb}{1, 0, 1}
\definecolor{errorcolor}{rgb}{1, 0, 0}

\usepackage{alltt}
\usepackage[T1]{fontenc}
\usepackage[latin9]{inputenc}
\usepackage{geometry}
\usepackage{array}
\usepackage{float}
\usepackage{multirow}
\usepackage{graphicx}
\usepackage[authoryear]{natbib}

\makeatletter

\providecommand{\tabularnewline}{\\}

\usepackage{times}
\usepackage{graphics,natbib}

\makeatother
\IfFileExists{upquote.sty}{\usepackage{upquote}}{}
\begin{document}

\title{Analyzing trends in precipitation patterns using Hidden Markov model
stochastic weather generators}
\author{Christopher Paciorek\\
Department of Statistics, UC Berkeley}

\maketitle

\section*{Abstract}

We develop a flexible spline-based Bayesian hidden Markov model stochastic
weather generator to statistically model daily precipitation over
time by season at individual locations. The model naturally accounts
for missing data (considered missing at random), avoiding potential
sensitivity from systematic missingness patterns or from using arbitrary
cutoffs to deal with missingness when computing metrics on daily precipitation
data. The fitted model can then be used for inference about trends
in arbitrary measures of precipitation behavior, either by multiple
imputation of the missing data followed by frequentist analysis or
by simulation from the Bayesian posterior predictive distribution.
We show that the model fits the data well, including a variety of
multi-day characteristics, indicating fidelity to the autocorrelation
structure of the data. Using three stations from the western United
States, we develop case studies in which we assess trends in various
aspects of precipitation (such as dry spell length and precipitation
intensity), finding only limited evidence of trends in certain seasons
based on the use of Sen's slope as a nonparametric measure of trend.
In future work, we plan to apply the method to the complete set of
GHCN stations in selected regions to systematically assess the evidence
for trends.\pagebreak{}

\section{Introduction}

Changes in precipitation are a critical component of potential impacts
from climate change, via a variety of specific mechanisms, including
flooding, drought, agricultural productivity, and effects on natural
ecosystems \citep{nca4-II,ipcc2022-wg2}. Much research has focused
on assessing the evidence for trends in overall precipitation and
in extreme precipitation \citep{nca4-I,ipcc2021-wg1}, sometimes using
statistical extreme value analysis when focusing on very extreme precipitation
\citep[e.g.,][]{westra2013global,risser2019detected}. There is also
an emphasis on formal \textquotedbl detection and attribution\textquotedbl{}
analyses in the climate science literature, which attempt to lend
statistical robustness to statements about the causal relationship
of climate forcing variables (in particular greenhouse gases) with
precipitation \citep[e.g.,][]{zhang2007detection}. Another thread
in the literature focuses on attributing individual extreme events,
many of which are at least partly focused on precipitation \citep{nas2016}. 

While extreme individual events (often daily extremes) are of course
critically important, many questions about imacts and the fundamental
climate science for understanding the mechanisms of changes relate
to the entire pattern of precipitation, over the course of multiple
days or much longer time periods. For such analyses, standard extreme
value analyses are not as relevant or provide limited information
because of limited data availability inherent in using time periods
longer than a day. For example, the near-failure of the Oroville Dam
in California occurred during a winter with high overall precipitation
but for which no individual days or even events were particularly
extreme (O'Brien et al., in prep.). As a second example, a critical
question for agriculture is whether the frequency of droughts is changing
with climate change. \citet{zhang2021five} report trends in mean
dry interval length potentially indicative of increasing drought in
the southwestern United States. More generally, a primary hypothesis
about the impact of global warming on precipitation is one of intensification
of dry and wet events, as a warming atmosphere is able to hold more
water \citep*{zhang2021increasing} 

In light of this, there is a need for statistical methods that go
beyond analysing aggregate measures such as means or extremes (either
annual or seasonal). One can develop metrics for phenomena of interest,
such as the mean dry spell length (DSL) metric used in \citet{zhang2021five}
and various metrics considered in O'Brien et al. (in prep.), and then
carry out statistical analysis of trends on the metrics, but this
has several shortcomings. First, one runs into multiple testing issues
if considering multiple metrics, which will almost always be the case
given the arbitrariness of any metric and the need for sensitivity
analyses. Second, the use of complementary metrics to characterize
the system (e.g., mean dry spell length and mean precipitation during
wet events) may be of interest, and again multiple testing considerations
would come into play. Perhaps most importantly, many of the metrics
of interest cannot be computed in the face of missing data, which
are common in observational data products. For example, one cannot
compute the length of a dry interval in the face of a missing observation,
with no real alternative to handle this apart from throwing out data
(potentially entire seasons or years) contaminated by missingness.
This is much more concerning than when computing mean precipitation,
where an assumption that the missingness is non-informative can allow
one to still compute a usable summary statistic. This missingness
leads to omission of usable data (e.g., from years with only some
data) and inability to assess trends further back in time as missingness
becomes more common. This work seeks to develop a statistical method
that can be used in this situation and to assess whether interesting
trends in precipitation patterns can be detected from individual stations
in light of the known low signal-to-noise ratio in station-level precipitation
data \citep{chandler2014uncertainty,nca4-I} (and see references in
\citealp{risser2019detected}).

We propose to use statistical models fit to daily data to characterize
the patterns of precipitation and trends over time within a single
model that can then be used for self-consistent inference about any
metric of interest. The approach considers a metric to be a summary
statistic functional of the joint distribution (over days) of precipitation,
where the joint distribution accounts for the underlying temporal
dependence structure of precipitation. Stochastic weather generators
(SWGs) aim to account for the complicated temporal dependence of precipitation
and hold promise for this sort of analysis, but the use of SWGs has
focused on just that, serving to generate realistic weather time series
\citep{wilks1999weather}, including for downscaling, rather than
for carrying out inference about temporal changes. Various forms of
SWGs have been proposed (see summaries in \citealp{chandler2014uncertainty}
and \citealp{ailliot2015stochastic}), including hidden Markov models
(HMMs) \citep[e.g.,][among many others]{holsclaw2017bayesian,stoner2020advanced},
generalized linear models (GLMs) \citep[e.g.,][]{chandler2002analysis,yang2005spatial},
and regional climate models. This work uses HMMs to characterize daily
precipitation patterns, building in spline-based functions to account
longer-term trends, including potential effects of climate change
as well as modes of internal variability. The approach naturally handles
missing observations via the usual HMM forward algorithm, with simple
imputation of missing values using standard statistical methods.

We fit our HMM-based SWG using a Bayesian approach and use the fitted
model as a statistical model to assess potential trends in two ways.
First, we treat the problem as a missing data problem amenable to
multiple imputation-based frequentist analysis. We make multiple imputations
of the missing precipitation values using the fitted HMM and then
use simple analyses of trend in any metric of interest, accounting
for the imputation uncertainty. Second, we treat the problem in a
fully Bayesian fashion, using the posterior of the fitted HMM to generate
from the posterior predictive distribution of the metric of interest
by simulating many time series of precipitation. The former has the
advantage of relying on the data as much as possible and the model
only for imputation, while the latter has the advantage of giving
coherent posterior inference for any quantities of interest. Via either
approach, we can then carry out trend analysis (e.g., using Sen's
slope \citep{sen1968estimates} or linear regression) to estimate
trends with uncertainty. We regard these trend estimates as useful
first-order summaries of temporal variation, rather than believing
that trends truly are monotonic or linear, in the spirit of \citet{woody2021model}. 

The modeling approach is purely empirical in that it doesn't rely
on extra information (such as other datasets, reanalysis, or even
observations of quantities other than precipitation). This is beneficial
for focused empirical work that seeks to report the evidence for trends
based only on observations, and disadvantageous for other purposes
as it leverages limited information and does not take account of the
larger weather context about which much is known, particularly more
recently in time with the availability of satellite data. The fitted
model could also be used as a stochastic weather generator, though
because it is unconnected to any forcing variables (i.e., covariates)
and fit at individual locations, it has limitations in terms of when
it would be appropriate to use. 

We apply the methodology to several case studies related to specific
scientific questions about potential trends in precipitation patterns,
using data from the Global Historical Climatology Network (GHCN),
which have been processed using various quality assurance checks \citep{menne2012global}.
We consider potential trends over the periods 1920-2021, 1950-2021,
and 1980-2021 out of scientific interest in trends over intervals
of varying length, as well as diminishing data prior to the 1920s,
although we fit the HMM for 1900-2021 to include a boundary period.

The first case study is of drought in the southwestern U.S., motivated
by the analysis of \citet{zhang2021five}. We chose a station with
nearly complete data in Winslow, Arizona (AZ) that was also used in
\citet{zhang2021five}. For this station, we focus on mean dry spell
length (DSL) in each of the four standard seasons (DJF, MAM, JJA,
SON), as a metric of meteorological drought. We also consider precipitation
intensity, defined as the mean precipitation on days with precipitation.

The second case study is of prolonged dry spells and precipitation
variability in California during the wet season (roughly October through
April). For this we chose a station with nearly complete data in Berkeley,
California (CA), focusing on mean dry spell length and precipitation
intensity for each of three seasons (DJF, MAM, and SON, omitting JJA
as essentially no precipitation falls in the area during the summer
months). In recent years, California has experienced serious drought
conditions, interspersed with wet years \citep{wang2017california,swain2018increasing}.
In some of these years, there have been long periods during the wet
season without precipitation, associated with a persistent high pressure
system in the Pacific Ocean that blocks precipitation events (such
as atmospheric rivers) \citep{wang2014probable,swain2018increasing},
including a very long dry spell this past winter during January to
March 2022. Our analysis focuses on assessing the evidence for trends
in dry spell length and precipitation intensity as measures of such
blocking behavior and the possibility of more pronounced precipitation
variability expected with climate change \citep{wood2021changes}.

The third case study is of precipitation patterns in the mountains
of California, focused on the Feather River basin northeast of Sacramento.
In the 2017 wet season, the Oroville Dam, a large dam that captures
the flow of the Feather River, was severely damaged \citep{kasler2017as}
because of high precipitation over a period of weeks. For this analysis
we chose a station centrally located in the basin and with a nearly
complete record in Quincy California. O'Brien et al. (in prep.) concluded
that while no individual precipitation events were particularly extreme,
there was a collection of many moderately-intense events during the
time preceding the damage to the dam. Following O'Brien, our analysis
focuses on assessing evidence for trend in the number of events, the
average precipitation during the events, and maximum 40-day precip
(a metric suggested in \citet{swain2018increasing} in part because
of the historic 1861-1862 Sacramento flooding associated with extremely
high precipitation over a similarly-lengthy time period). Given the
hydrologic context, for this case study, we consider the full wet
season defined as November through April. 

In Section \ref{sec:Methods} we describe the Bayesian HMM and MCMC
fitting approach, as well as our approach to model selection and assessment.
In Section \ref{sec:Results}, we assess the fitted model and then
apply the methodology to the three case studies. 

\section{Methods\label{sec:Methods}}

Our model is an extension of the HMM developed by \citet{stoner2020advanced}
to model hourly precipitation. We modify the model to use with daily
precipitation data, with flexible spline-based terms to capture variation
over days within a season (\emph{seasonal} terms) and variation over
years (\emph{yearly} terms). We fit the model separately for each
season (generally DJF, MAM, JJA, SON), as the atmospheric drivers
of precipitation vary widely over the course of a year (e.g., frontal
systems in winter and convective systems in summer), and we have no
reason to expect any time trends to be similar across seasons, related
to both dynamics and thermodynamics. This approach allows for there
to be different long-term trends for different seasons without requiring
specification of a complicated season-year interaction (discussed
further below).

\subsection{Hidden Markov model specification}

As in \citet{stoner2020advanced}, our primary model uses three clone
dry states and two wet states. The clone dry states allow for realistic
multi-day dry spells, which would not be captured by a single dry
state with a geometrically distributed number of days spent in the
state \citep{stoner2020advanced}. The term \emph{clone} reflects
that all three of the dry states have the same conditional distribution
for rainfall. Transitions between the dry states are not allowed,
producing a mixture of geometric distributions for the time spent
in the dry states. For the distribution over states on the first day
of a season within each year, we use a Dirichlet distribution. For
simplicity this distribution is assumed to not change over time.

\subsubsection{Transition probabilities}

The elements of the HMM state transition matrix for the five states
are determined by $p_{jh}=Pr(z_{st}=h|z_{s-1,t}=j$) for $h,j\in\{1,2,3,4,5\}$,
where $s$ indexes the day within a season and $t$ indexes year,
and where the two wet states are $h\in\{4,5\}$. The full matrix,
$P=\{p_{jh}\}$, is 
\[
\left(\begin{array}{ccccc}
p_{D_{1}} & 0 & 0 & q_{1} & *\\
0 & p_{D_{2}} & 0 & q_{2} & *\\
0 & 0 & p_{D_{3}} & q_{3} & *\\
u_{11} & u_{12} & * & p_{W_{1}} & r_{12}\\
u_{21} & u_{22} & * & r_{21} & p_{W_{2}}
\end{array}\right),
\]
where $*$ indicates a entry computed such that each row sums to 1
and 
\begin{eqnarray*}
q_{1} & = & (1-p_{D_{1}})p_{DW_{1}}\\
q_{2} & = & (1-p_{D_{2}})p_{DW_{1}}\\
q_{3} & = & (1-p_{D_{3}})p_{DW_{1}}\\
r_{12} & = & (1-p_{W_{1}})p_{W_{12}}\\
r_{21} & = & (1-p_{W_{2}})p_{W_{21}}\\
u_{11} & = & (1-p_{W_{1}}-r_{12})p_{WD_{1}}\\
u_{21} & = & (1-p_{W_{2}}-r_{21})p_{WD_{1}}\\
u_{12} & = & (1-p_{W_{1}}-r_{12}-u_{11})p_{WD_{2}}\\
u_{22} & = & (1-p_{W_{2}}-r_{21}-u_{21})p_{WD_{2}}.
\end{eqnarray*}

Note that there are certain constraints built into the matrix, generally
following \citet{stoner2020advanced}, to reduce the number of parameters,
given this is already a flexible formulation with multiple dry and
wet states. In particular, conditioning on the probability of transitioning
from a dry state to a wet state, the probability of transitioning
to a given wet state does not depend on the dry state being transitioned
from. Similarly, conditioning on the transition from a wet state to
a dry state, the probabilities of transitioning to a given clone dry
state do not differ based on the wet state being transitioned from.

To account for variation in time, both long-term variation from internal
variability and long-term trend potentially related to climate change,
as well as variation over the course of a season, we use an additive
representation with two spline terms for $\mbox{logit}(v)$ where\\
 $v\in\{p_{D_{1}},p_{D_{2}},p_{D_{3}},p_{DW_{1}},p_{W_{1}},p_{W_{2}},p_{W_{12}},p_{W_{21}},p_{WD_{1}},p_{WD_{2}}\}$:
\[
\mbox{logit}(v_{st})=\beta_{0}^{v}+X(s)\beta_{s}^{v}+X(t)\beta_{t}^{v}.
\]
Consistent with the notion of clone states, the basis coefficients
for the three dry states are taken to be equal: $\beta_{s}^{p_{D_{1}}}=\beta_{s}^{p_{D_{2}}}=\beta_{s}^{p_{D_{3}}}$
and $\beta_{t}^{p_{D_{1}}}=\beta_{t}^{p_{D_{2}}}=\beta_{t}^{p_{D_{3}}}$. 

The basis matrices are computed using the \emph{jagam} function from
the \emph{mgcv} package \citep{wood2017generalized}, using the default
thin splate regression spline with a dimension of $K=20$, chosen
to be able to account for variation at the scale of weeks/months (seasonal
terms) and decades (yearly terms), but not variability at the scale
of individual days of the year or individual years. The splines are
penalized in a Bayesian fashion using individual shrinkage priors
(based on reparameterizing the splines using the \emph{diagonalize=TRUE}
argument in \emph{jagam}):
\begin{eqnarray*}
\beta_{s,i}^{v} & \sim & \mathcal{N}(0,\sigma_{s}^{2,v}),\,i=1,\ldots,K-1\\
\beta_{t,i}^{v} & \sim & \mathcal{N}(0,\sigma_{t}^{2,v}),\,i=1,\ldots,K-1
\end{eqnarray*}
with noninformative priors for $\beta_{s,K}^{v}$ and $\beta_{t,K}^{v}$,
which are associated with linear terms of season and year. For the
variance parameters, we use flat priors on the standard deviation
scale (Gelman 2006), specifically uniform on (0.001, 10) to prevent
the MCMC from wandering in non-identifiable parts of the space for
the variance components. For the intercept terms we use reasonably
non-informative priors restricted to avoid placing heavy weight on
extreme probabilities. 

To avoid additional complexity from another spline term, we assume
that the seasonal effect does not change across years. We consider
this a first-order approximation, as changes in seasonality may well
occur over many years, but we feel this is justified given the limited
power seen in detecting changes in the main effect of year (Section
\ref{subsec:Case-Studies}).

\subsubsection{Precipitation distribution}

There has been extensive discussion in the climate science and statistics
literature on appropriate distributions for precipitation. \citet{furrer2008improving}
find that gamma distributions have too light a tail and cannot capture
the right tail of the distribution accurately. \citet{stoner2020advanced}
use a generalized Pareto distribution (GPD) for non-zero precipitation
in each HMM state, while \citet{naveau2016modeling} develop a generalization
of the GPD. In contrast \citet{martinez2019precipitation} argue for
the use of gamma distribution. We considered using both the gamma
distribution and the generalized Pareto distribution. Since we found
the gamma distribution to perform best (Section \ref{subsec:Model-comparison}),
we next present that as the primary model.

The model uses a mixture of a point mass at zero and a gamma distribution
for precipitation conditional on state, $p(R_{st}|z_{st})$. The three
dry states are considered as a single state, with a single precipitation
distribution, while each of the two wet states has their own precipitation
distribution. Thus for each of these three distributions, $k\in\{1,4,5\}$,
we have:
\[
p(R_{st}=r_{st}|z_{st}=k)=\pi^{k}+(1-\pi^{k})Ga(\xi_{st}^{k},\sigma_{st}^{k}),
\]
where $\xi$ is the shape parameter and $\sigma$ the scale parameter.
For the log of $\xi_{st}^{k}$ and $\sigma_{st}^{k}$, we use the
same spline representations of season and year as described above
for the logit of the probabilities used to construct the HMM state
transition matrix. The probability of no daily precipitation, $\pi^{k}$,
is assumed not to vary over season or year, to help reduce non-identifiability
from the HMM state transition probabilities trading off with the probability
of precipitation given state. To help enforce that the dry state would
generally have low precipitation, we assume there is no variation
over season or year for the dry state ($\xi_{st}^{1}=\xi^{1}$ and
$\sigma_{st}^{1}=\sigma^{1}$). 

As an alternative model, we also considered the GPD in place of the
gamma distribution above, as in \citet{stoner2020advanced}, with
$\xi_{st}^{k}$ and the log of $\sigma_{st}^{k}$ being spline-based
representations of season and year for the shape and (log) scale parameters
of the GPD.

\paragraph{Rounding}

The GHCN data are rounded to the nearest .01 inch. To account for
this, we create a likelihood for the discretely-observed values by
integrating the underlying gamma distribution (or GPD) over plus or
minus .005 inches of the reported value. That said, the degree of
rounding is not so discretized that one would necessarily expect that
the usual assumption of continuous values would pose a problem. For
some stations (e.g., Quincy, CA but not Berkeley, CA), there appears
to be some additional rounding to the nearest 0.1 inches or (in some
cases 0.05 inches) for some of the values, with more such apparent
rounding in earlier years. We do not account for the rounding to coarser
increments as this would necessitate modeling to account for the unknown
fraction of observations rounded to different degrees.

\subsubsection{Identifiability}

Models with unknown states, such as HMMs, are prone to identifiability
issues, which can affect MCMC convergence and mixing. While we are
not able to completely avoid this, we impose some constraints to reduce
the problem, in addition to the already-specified constraints built
into the model. Specifically, to enforce the meaning of dry versus
wet states, we constrain $\pi^{1}>\pi^{4}>\pi^{5}$, and we enforce
that the mean of the gamma distribution (or median of the GPD) are
increasing over the states (from dry to wet state 1 to wet state 2)
separately for every season-year value during the MCMC fitting. We
also constrain $p_{st}^{WD_{1}}>0.4$ and $p_{st}^{WD_{2}}>0.4$ for
each value of $s$ and $t$ to encourage that the first dry state
be entered most frequently and the third dry state least frequently.
Finally, in early runs, we found that the parameters of the second
wet state could wander off into parts of the parameter space unconstrained
by the data and produce unrealistically large precipitation values
when simulating from the posterior predictive distribution (simulating
precipitation in the thousands or even millions of cm), generally
when the probability of entering the second wet state was low (e.g.,
during seasons that are generally dry in a given location). To avoid
this, we constrain the mean of the gamma distribution (or median of
the GPD) at every season-year time point to be less than the maximum
daily precipitation over all years for a given location-season.

In our MCMC assessments, we examined trace plots and generally found
reasonable mixing for the individual parameters in the model, but
with some indications of non-identifiability, for example with the
shape and scale parameters of the gamma distributions for the two
wet states transitioning between different levels that correspond
to changes in some of the probabilities in the transition matrix.
This is not surprising given the multiple dry (clone) and wet states
of the HMM. Given the inherent non-identifiability, we focused assessment
of mixing on various summary statistics of the posterior predictive
distributions (Section \ref{subsec:Fitting}).

\subsubsection{Seasonal modeling}

Our initial model attempted to represent all four seasons in a single
model. It used three spline terms to capture temporal trends across
day of season, year, and the interaction of the two. This required
careful consideration of the spline basis terms to avoid strong posterior
correlations \citep{wood2017generalized}, and even with this, we
found that MCMC mixing was rather worse than the single-season models.
For simplicity, we chose to stratify by season as discussed above,
rather than proceeding with the full season-year model. However, we
note that the mixing difficulties revealed themselves in terms of
non-identifiability given the multiple dry and wet states, as discussed
above, and if one focuses on posterior predictive quantities reflecting
the model's characterization of the observable precipitation patterns,
the full model might be worth further consideration. 

\subsection{Fitting\label{subsec:Fitting}}

For Bayesian estimation using Markov chain Monte Carlo (MCMC), one
can choose whether to estimate latent states or to integrate over
them, using a variety of techniques. For HMMs, the integration is
straightforward using the standard forward algorithm, which we use.
We assume missing data as having their missingness unrelated to the
unknown value, but of course related to the time point (day of season
and year), so considered to be missing at random (MAR), accounted
for by the season and year spline terms.

We use NIMBLE version 0.12.2 \citep{devalpine2017programming,nimble_development_team_2022_6317014}
to carry out the MCMC, implementing the forward algorithm as a user-defined
function. The bulk of the computation in the MCMC involves calculation
of the precipitation density for each observation given each potential
state and the forward algorithm computations. We also note that the
discretized likelihood incurs the cost of multiple gamma CDF calculations,
which are costly. We are careful to only recalculate the density values
for each observation under each potential state when the parameters
of the precipitation distribution (and not the parameters of the transition
matrix) change. For missing values, we impute precipitation using
the forward filtering backward sampling (FFBS) algorithm, implemented
as a user-defined sampler in NIMBLE. This imputes from the full posterior
distribution for the states at all times and then draws from the precipitation
distribution given the imputed state. Note that one could draw from
the forward algorithm's probability distribution over the states,
but this would ignore the information from the future that is relevant
for prediction, although the impact of this would be lessened for
longer strings of sequential missing values. 

For each location-season of interest, we fit the model to all data
in 1900-2021 (starting with winter of the year 1900, which includes
months from 1899) and concluding with the fall season of 2021). We
ran five MCMC chains of 20,000 iterations each (each chain taking
about one day on a single CPU core), taking a conservative burn-in
period of 5,000 iterations for each chain, and thinning by 10 to reduce
storage and computation, producing a total of 7,500 posterior samples.
We use NIMBLE's adaptive Metropolis samplers for each scalar parameter.
Initial exploration of other samplers (blocking, slice sampling, etc.)
did not indicate better performace from other approaches. For posterior
predictive quantities, we simulated one full time series for each
of the 7,500 posterior samples. 

Given the aforementioned indications of non-identifiability, we focus
on posterior predictive quantities from the model that characterize
the climatology that the model produces for the period 1920-2021.
These include
\begin{itemize}
\item several values of the precipitation CDF (greater than 3 mm, 10 mm,
20 mm) by year and by day of season;
\item transition probabilities between dry and wet days (using a cutoff
of 3 mm) by year;
\item mean dry spell length and precipitation intensity by year; and
\item Sen's slope for dry spell length and precipitation intensity, as an
aggregate summary.
\end{itemize}
Using posterior predictive time series simulations, we compute the
$\hat{R}$ and ESS as suggested in \citet{vehtari2021rank} and available
in the RStan R package version 2.21.2 \citep{stan_development_team}.
Given the large number of diagnostic quantities (e.g., one per year
for each location-season pair in many cases), we report summaries
of the distribution (over the various years or days of season) of
these diagnostics for the various quantities for each of the eight
location-season pairs of the case studies. Table \ref{tab:Median-(99th-percentile)}
(Appendix) shows percentiles of $\hat{R}$ for the eight location-seasons.
In essentially all cases, mixing seems reasonable. Table \ref{tab:MCMC-diagnostics-(and}
shows $\hat{R}$, ESS, and a measure of ESS focused on tail quantities
\citep{vehtari2021rank} for Sen's slope. In most cases, mixing seems
reasonable, but for JJA for Winslow for mean dry spell length, there
is indication of concern (and perhaps for DJF for Berkeley for both
dry spell length and precipitation intensity and JJA for Winslow for
precipitation intensity). This is somewhat surprising given the diagnostics
for the yearly mean dry spell length and precipitation intensity (Table
\ref{tab:Median-(99th-percentile)} (Appendix), columns 11-12) do
not indicate any issues. It appears that Sen's slope may be be sensitive
to very subtle changes in the yearly values that are present in the
posterior samples, perhaps related to label-switching/non-identifiability
of the HMM. Considering Sen's slope over later periods (1950-2021
and 1980-2021), the $\hat{R}$ values for mean dry spell length for
JJA for Winslow are smaller (1.036 and 1.010). In future work, we
plan to explore whether the use of Hamiltonian Monte Carlo improves
upon the MCMC sampling strategy used here.

\begin{table}
\caption{\label{tab:MCMC-diagnostics-(and}MCMC diagnostics ($\hat{R}$ and
bulk ESS and tail ESS) for Sen's slope (for 1920-2021) for mean dry
spell length and precipitation intensity for the eight location-seasons.}

\begin{tabular}{|c|c|c|c|c|c|c|}
\hline 
 &
\multicolumn{3}{c|}{mean dry spell length} &
\multicolumn{3}{c|}{precipitation intensity}\tabularnewline
\hline 
 &
$\hat{R}$ &
ESS, bulk &
ESS, tail &
$\hat{R}$ &
ESS, bulk &
ESS, tail\tabularnewline
\hline 
\hline 
Winslow AZ, DJF &
1.012 &
1114 &
3515 &
1.005 &
2545 &
6101\tabularnewline
\hline 
Winslow AZ, MAM &
1.000 &
5982 &
5744 &
1.000 &
4978 &
6733\tabularnewline
\hline 
Winslow AZ, JJA &
1.089 &
35 &
121 &
1.015 &
361 &
5750\tabularnewline
\hline 
Winslow AZ, SON &
1.000 &
5075 &
6364 &
1.000 &
5410 &
6492\tabularnewline
\hline 
Berkeley CA, DJF &
1.015 &
424 &
5746 &
1.024 &
184 &
1955\tabularnewline
\hline 
Berkeley CA, MAM &
1.006 &
3605 &
6718 &
1.011 &
1351 &
5623\tabularnewline
\hline 
Berkeley CA, SON &
1.000 &
6507 &
6928 &
1.001 &
5587 &
6768\tabularnewline
\hline 
Quincy, wet (Nov-Apr) &
1.001 &
4694 &
6548 &
1.001 &
5115 &
7144\tabularnewline
\hline 
\end{tabular}
\end{table}

\subsection{Model selection}

We do not carry out extensive model selection, as we have constructed
a flexible model that nests simpler models within it. The use of separate
variances for the spline term coefficients, estimated as part of the
fitting process, allows for shrinkage of the day-of-season- and year-varying
effects for the various states, building in the ability for the fitting
process to estimate simpler models when sufficient to fit the data
well, as would be favored by the usual inherent Bayesian complexity
penalty \citep{jefferys1992ockham,mackay2003information}. With regard
to the number of states, if fewer states are needed, the fitting could
estimate the transition probability into an unneeded state to be near
zero, thus allowing the spline terms for that state to shrink to having
no seasonal or yearly variation.

That said, we do carry out two types of model comparison for a subset
of location-season pairs (DJF and JJA for Winslow, DJF for Berkeley,
and November-April for Quincy, ). First we compare the gamma distribution
and GPD for precipitation. Second, we compare the full model to a
model with no yearly trend terms as a guard against overfitting. We
use both a held-out test set and WAIC. For the assessment on the test
set, we hold out individual years of data at a time (every tenth year
for the years 1910 through 2020) and assess one-day ahead predictions
and full-year predictions. I.e., the one-day ahead cross-validation
calculates the likelihood of the held-out observation for a given
day, with the prediction conditioned on all the data before that day
(including the earlier held-out data). The full-year predictions condition
only on the training data, using the forward algorithm to calculate
the predictive distribution for each held-out observation. In addition,
to compare the gamma and GPD-based models, we examine Q-Q plots, as
a key distinction between these models is the right tail of the distribution. 

Note that WAIC is a pointwise measure, so one must choose the unit
of observation \citep{vehtari2017practical}, and for temporal data,
it is not clear that treating each day as an observation is appropriate
any more than it would be for cross-validation. Instead we choose
to treat each year as a multivariate observation, which is easily
done when requesting WAIC calculation in NIMBLE \citep{hug2021numerically}.
Also note that for logistical simplicity and to be able to compare
held-out and WAIC results from the same model fits, we calculate WAIC
in conjunction with the hold-out test set analysis, so the training
data for WAIC are not the full dataset (i.e., every tenth year is
excluded).

\subsection{Model assessment}

Given the a best model based on the model selection process, we assess
the fit of the model by comparing summary statistics using the posterior
predictive distribution, using the 7,500 simulated precipitation time
series and masking the simulated series to match the missingness in
the actual data (with one exception noted later). We consider the
following summary statistics:
\begin{itemize}
\item total season precipitation as a function of year;
\item probability of daily precipitation as a function, marginally, of year
and season;
\item precipitation intensity (mean precipitation on days with precipitation
greater than 3 mm) as a function, marginally, of year and season;
\item conditional distributions for (discretized) precipitation given (discretized)
previous day precipitation as a function, marginally, of year and
season;
\item Q-Q plots for daily precipitation, three-day precipitation and ten-day
precipitation over all years and seasons, where three- and ten-day
precipitation are calculated using all possible (overlapping) moving
blocks, thereby capturing the full precipitation of entire events;
and
\item mean dry spell length as a function of year, and Q-Q plots for dry
spell length, where precipitation less than 3 mm is considered a dry
day, and spells are truncated by the beginning and end of a season.
\end{itemize}
Some of these diagnostics target various marginal distributions and
don't assess model performance relative to some of the multi-day phenomena
of interest, while others directly address multi-day fidelity.

Note that imposing the same missingness pattern on the simulated values
as is present in the actual data allows for comparison of the empirical
and the simulated but also means that the diagnostic plots should
not be interpreted as showing the within-season or yearly patterns
directly because the data are not missing completely at random.

\subsection{Trend analysis}

We use an imputation-based frequentist analysis as well as a fully
Bayesian approach to trend assessment. Given the widespread use of
Sen's slope in climate science, we focus on that statistic here, but
others (such as other forms of robust regression or simple linear
regression) could be easily substituted. We consider three time periods:
1920-2021, 1950-2021, and 1980-2021.

\subsubsection{Imputation-based analysis}

We take a frequentist-based multiple imputation approach, using the
7,500 posterior imputations for the missing observations. For each
imputation, we use the now-complete data to estimate Sen's slope.
We then use Rubin's rules for combining the results (after a continuity
correction \citep{helsel2006regional} applied to each Mann-Kendall
S statistic), with the p-value computed using the z-statistic for
the mean of the corrected S statistics.

\subsubsection{Model-based analysis}

We take a fully Bayesian predictive approach, using the 7,500 posterior
predictive time series simulations. We consider trend estimates to
be summary statistics, rather than parameters (or direct functionals
of parameters) of the model, in the spirit of \citet{woody2021model}.
The fully Bayesian approach has the benefit, compared to the imputation-based
approch, of providing a Bayesian joint posterior distribution over
any functionals of interest. It has the downside of relying fully
on the HMM posterior predictive distribution being an adequate representation
of the aspects of precipitation affecting the metric of interest.
While our diagnostics indicate good fits (Section \ref{subsec:Model-assessment}),
one may be cautious about trends in extremes, such as extreme daily
values or multi-day events such as many-day precipitation or very
long dry or wet spells, for which one might be concerned that a first-order
HMM with gamma distributions for precipitation (even one with multiple
states) might not be adequate.

\subsection{Case studies}

Our analysis focuses on three case studies: (1) drought in the southwestern
United States using the Winslow AZ station; (2) winter (wet season)
precipitation variability in California using the Berkeley CA station;
and (3) winter (wet season) mountain preciipitation events using the
Quincy CA station. We use the GHCN data for these stations, omitting
observations failing any quality assurance checks, and setting trace
precipitation to zero.

For Winslow and Berkeley, we focus on mean dry spell length. Dry days
are considered to be those with less than three mm precipitation,
given evidence that such low amounts do not penetrate to the plant
rooting zone \citep{huxman2004precipitation}. We only consider days
within the season, so dry spells at the beginning and end of the season
are truncated. This simplification avoids having to make calculations
with days outside the season, keeping the inference focused on precipitation
pattens within the season, but with the caveat that the estimates
do not reflect the true lengths of dry spells. Note that this induces
a highly discrete distribution over mean DSL in seasons with few events,
with the mean DSL with one precipitation event equalling \textasciitilde 45
and with two precipitation events \textasciitilde 30. We also consider
precipitation intensity, defined as the mean precipitation on days
with precipitation, again defined based on the three mm cutoff. 

For Quincy, given the Oroville Dam event and interest in precipitation
events that led to the damage, we use metrics related to precipitation
events, reflecting intensity and frequency of the events, in particular
the number of wet spells in the wet season, mean precipitation of
the wet spells, and maximum precipitation of any 40-day period in
the season.

\section{Results\label{sec:Results}}

\subsection{Model comparison\label{subsec:Model-comparison}}

For our selected location-season pairs, model selection based on held-out
data and WAIC is equivocal with regard to using the gamma distribution
versus GPD (Table \ref{tab:Model-comparison-based}). However, for
certain GPD parameter values, the GPD can generate unrealistically
high precipitation (values in the hundreds, thousands, and even higher),
occurring when the shape parameter approaches one or greater. The
result is that the GPD-based simulated values have an overly heavy
far right tail. While Q-Q plots indicate that the overly heavy tail
only affects extreme precipitation (e.g., approximately the largest
0.2\% of values for Winslow AZ), it would raise the issue of what
to do with these values in an analysis. The suitability of the gamma
distribution even in the far right tail in light of results indicating
that a gamma distribution cannot capture the heavy tail of precipitation
\citep{furrer2008improving} likely relates in part to our mixture
over multiple states, which provides more flexibility in capturing
the underlying distribution. Another disadvantage of the GPD is that
in some MCMC samples, the right tail is bounded for some season-year
values (i.e., $\xi_{st}<0$) and one can find a small number of held-out
observations beyond the estimated bound and therefore deemed impossible
by the fitted distribution, complicating model comparison. Given the
similarity in the model comparison criteria and these disadvantages
of the GPD, we choose to use the gamma model.

With regard to trend, there is not clear evidence in favor of or against
the models with yearly trend in most cases, with the most evidence
for trend in JJA in Winslow. While there is not strong evidence for
overall trend, there is also little evidence that the full model is
overfitting. In light of this, we proceed with the full model because
doing so systematically allows us to avoid choosing between models
using what would necessarily be a somewhat arbitrary cutoff. The shrinkage
possible in the full model protects against overfitting and detecting
trends where there are none. As seen in our results, we find no evidence
of trends in many cases, suggesting that overfitted trends are not
a concern. An additional benefit of the full model is that the spline
terms can account for low-frequency variability, such as induced by
internal variability in the climate system (e.g., the recent multi-year
drought in the western U.S.).

\begin{table}
\caption{\label{tab:Model-comparison-based}Model comparison based on the negative
log-likelihood (NLL) of hold-out data for one-day ahead prediction
and full yearly prediction and WAIC (posterior standard deviation
of the negative log-likelihoods and pWAIC in parentheses). Lower values
indicate better fits. }

\begin{tabular}{|c|c|c|c|c|}
\hline 
 &
 &
NLL, one-day &
NLL, full-year &
WAIC (pWAIC)\tabularnewline
\hline 
\hline 
\multirow{3}{*}{Winslow (DJF)} &
gamma, no yearly trend &
926.1 (2.9) &
941.6 (2.4) &
14901 (53)\tabularnewline
\cline{2-5} \cline{3-5} \cline{4-5} \cline{5-5} 
 & gamma, yearly trend &
930.9 (3.8) &
943.5 (3.5) &
14860 (86)\tabularnewline
\cline{2-5} \cline{3-5} \cline{4-5} \cline{5-5} 
 & GPD, yearly trend &
926.5 (4.4) &
942.7 (4.3) &
14865 (55)\tabularnewline
\hline 
\multirow{3}{*}{Winslow (JJA)} &
gamma, no yearly trend &
1119.1 (2.5) &
1142.4 (2.2) &
21185 (61)\tabularnewline
\cline{2-5} \cline{3-5} \cline{4-5} \cline{5-5} 
 & gamma, yearly trend &
1114.1 (3.3) &
1137.2 (2.9) &
21145 (88)\tabularnewline
\cline{2-5} \cline{3-5} \cline{4-5} \cline{5-5} 
 & GPD, yearly trend &
1114.2 (4.0) &
1137.7 (3.9) &
21148 (74)\tabularnewline
\hline 
\multirow{3}{*}{Berkeley (DJF)} &
gamma, no yearly trend &
2323.9 (3.3) &
2430.1 (3.6) &
38284 (67)\tabularnewline
\cline{2-5} \cline{3-5} \cline{4-5} \cline{5-5} 
 & gamma, yearly trend &
2329.3 (4.1) &
2436.9 (5.2) &
38321 (143)\tabularnewline
\cline{2-5} \cline{3-5} \cline{4-5} \cline{5-5} 
 & GPD, yearly trend &
2331.3 (10.5) &
2439.0 (11.7) &
38299 (92)\tabularnewline
\hline 
\multirow{3}{*}{Quincy, wet (Nov-Apr)} &
gamma, no yearly trend &
3845.2 (2.6) &
4037.6 (4.0) &
76450 (60)\tabularnewline
\cline{2-5} \cline{3-5} \cline{4-5} \cline{5-5} 
 & gamma, yearly trend &
3849.5 (3.8) &
4038.5 (5.3) &
76383 (116)\tabularnewline
\cline{2-5} \cline{3-5} \cline{4-5} \cline{5-5} 
 & GPD, yearly trend &
3858.3 (20.6) &
4050.0 (23.1) &
76473 (110)\tabularnewline
\hline 
\end{tabular}

\end{table}

\subsection{Model assessment\label{subsec:Model-assessment}}

We next turn to assessing the fit of the chosen gamma model with yearly
trend terms based on the posterior predictive distribution over simulated
precipitation time series for the eight location-season pairs. 

\begin{figure}
\includegraphics[scale=0.75]{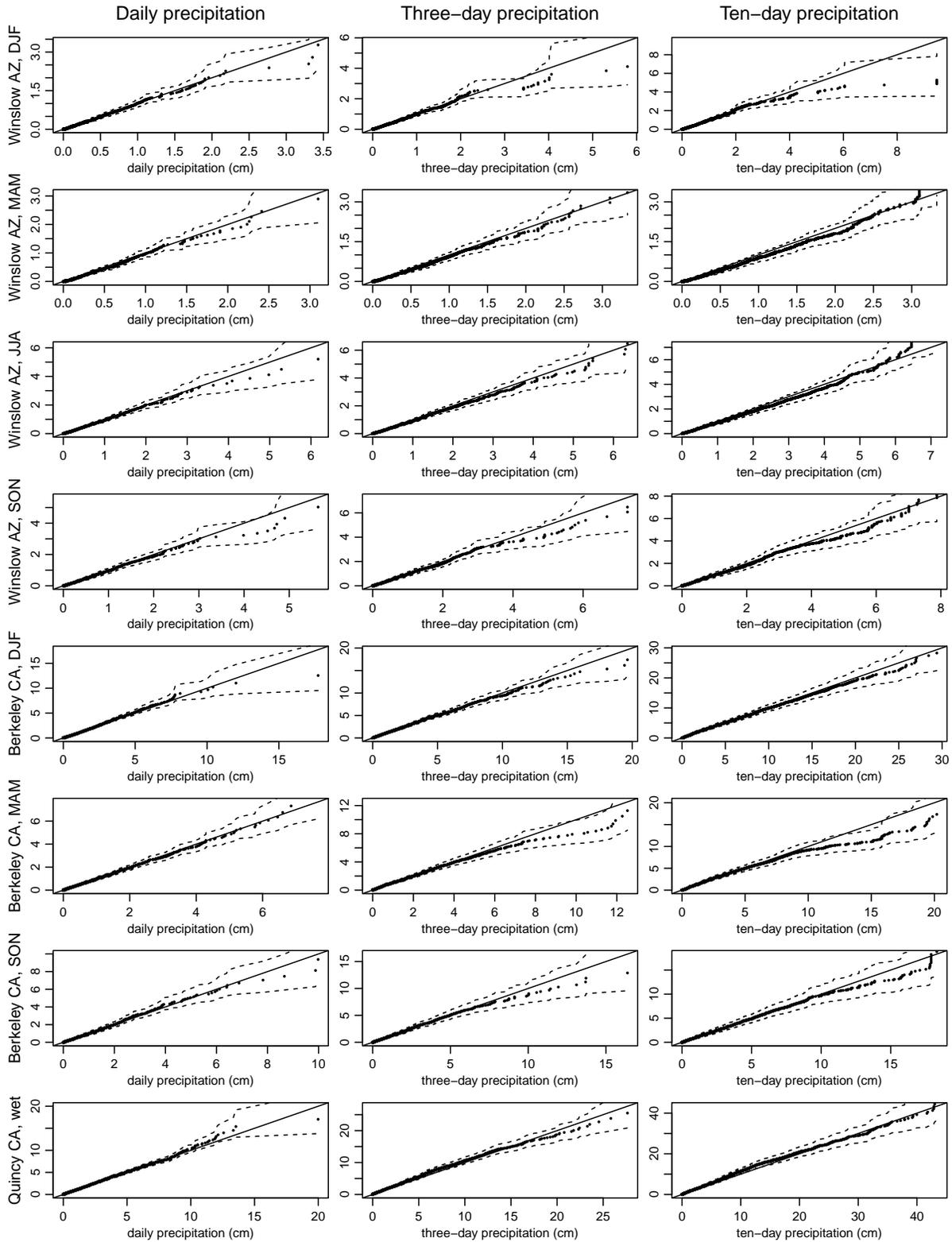}

\caption{\label{fig:Median-and-95=000025}Median and 95\% posterior predictive
distribution intervals for Q-Q comparisons of the simulated versus
observed daily (left column), three-day (middle), and ten-day (right)
precipitation (cm) for the eight location-season pairs (rows). The
median and 95\% intervals are taken over Q-Q comparisons of each of
the simulated time series against the observed values. If simulated
values are consistent with the observed values, we expect the points
(the median values) near the 1:1 line and the intervals to generally
cover the 1:1 line.}
\end{figure}

Fig. \ref{fig:Median-and-95=000025} shows quantile-quantile (Q-Q)
plots comparing the simulated and observed daily, three-day, and ten-day
precipitation. In general, the simulated distributions match the observed
distributions very well, with the main exception being that simulated
multi-day precipitation for MAM in Berkeley is too low for the most
extreme events.

\begin{figure}
\includegraphics[scale=0.75]{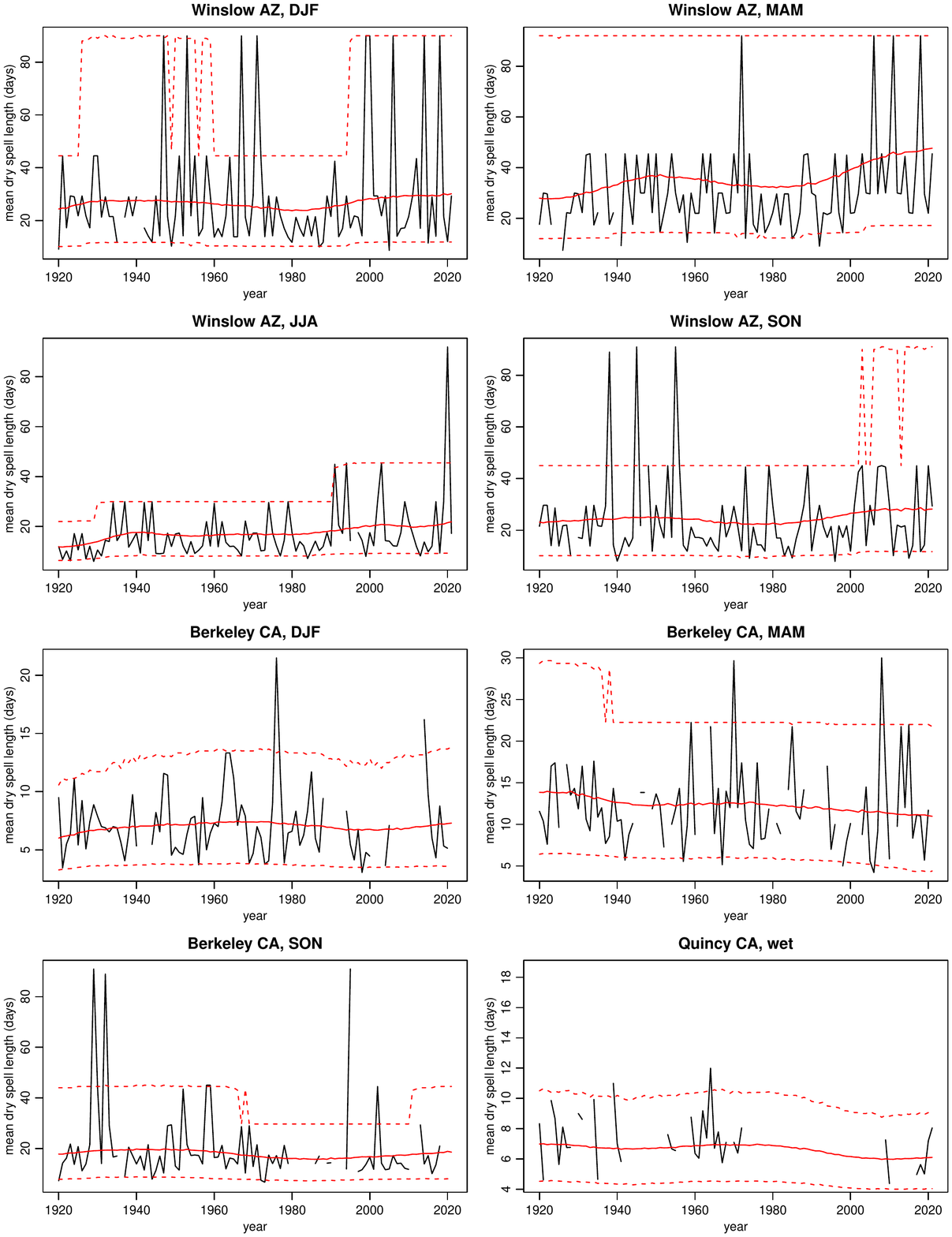}

\caption{\label{fig:Observed-(black)-versus}Observed (black) versus simulated
(red) mean dry spell length by year for the eight location-season
pairs, omitting observed values (but not simulated values, in order
to show predicted temporal variation) for years with any missing days.
Simulated values are the mean over the simulations, with 90\% posterior
predictive uncertainty bands. }
\end{figure}

Fig. \ref{fig:Observed-(black)-versus} shows the mean dry spell length
(DSL) by year for the observed and simulated values. The simulated
values match the observed values well, with simulation uncertainty
intervals doing a resonable job of covering the observations, albeit
with the uncertainty intervals appearing to overcover the observed
values. 

\begin{figure}
\includegraphics[scale=0.75]{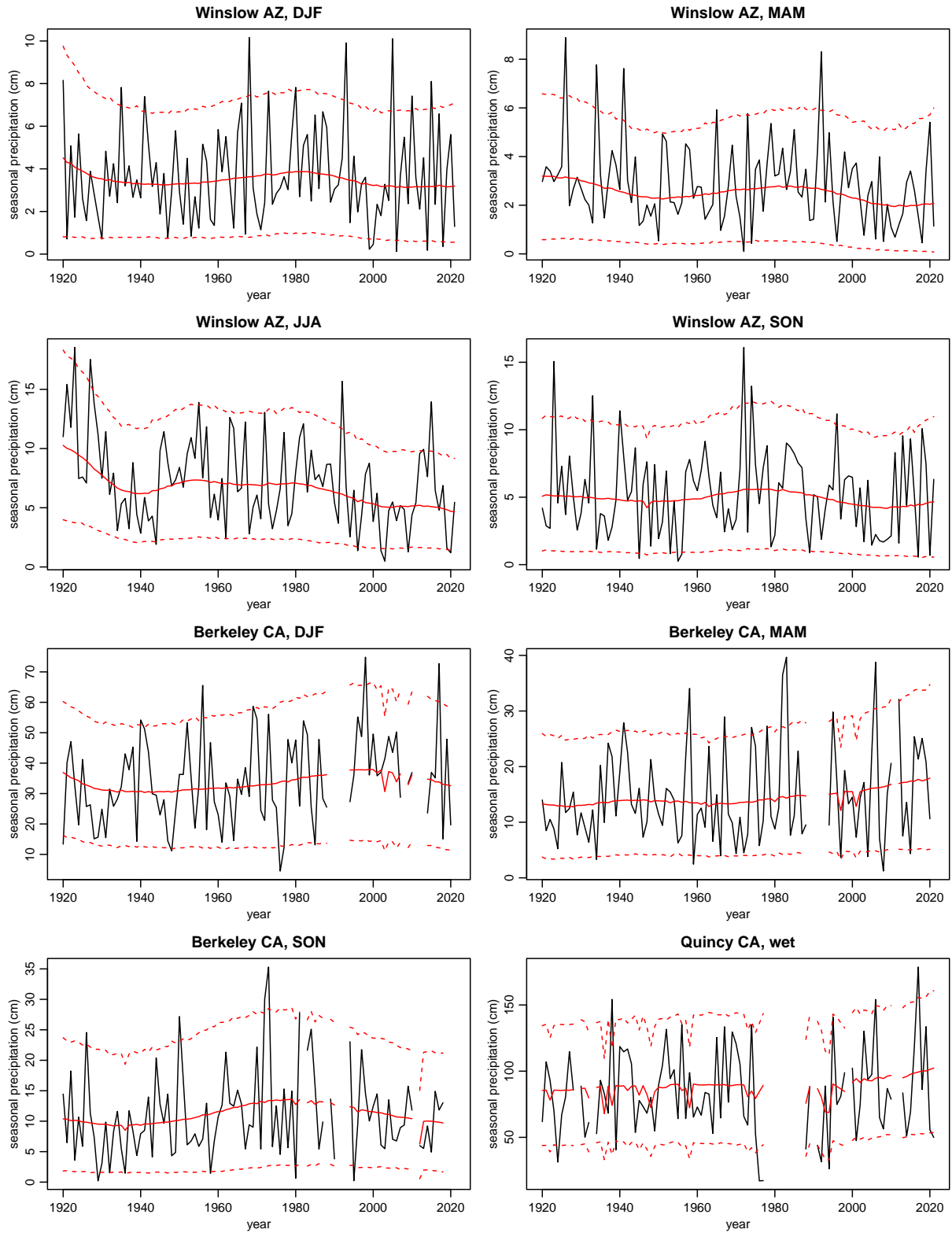}

\caption{\label{fig:diag_total}Observed (black) versus simulated (red) total
seasonal precipitation by year for the eight location-season pairs,
omitting years in which more than 25\% of observations were missing.
Simulated values are the mean over the simulations, with 90\% predictive
uncertainty bands.}
\end{figure}

\begin{figure}
\includegraphics[scale=0.75]{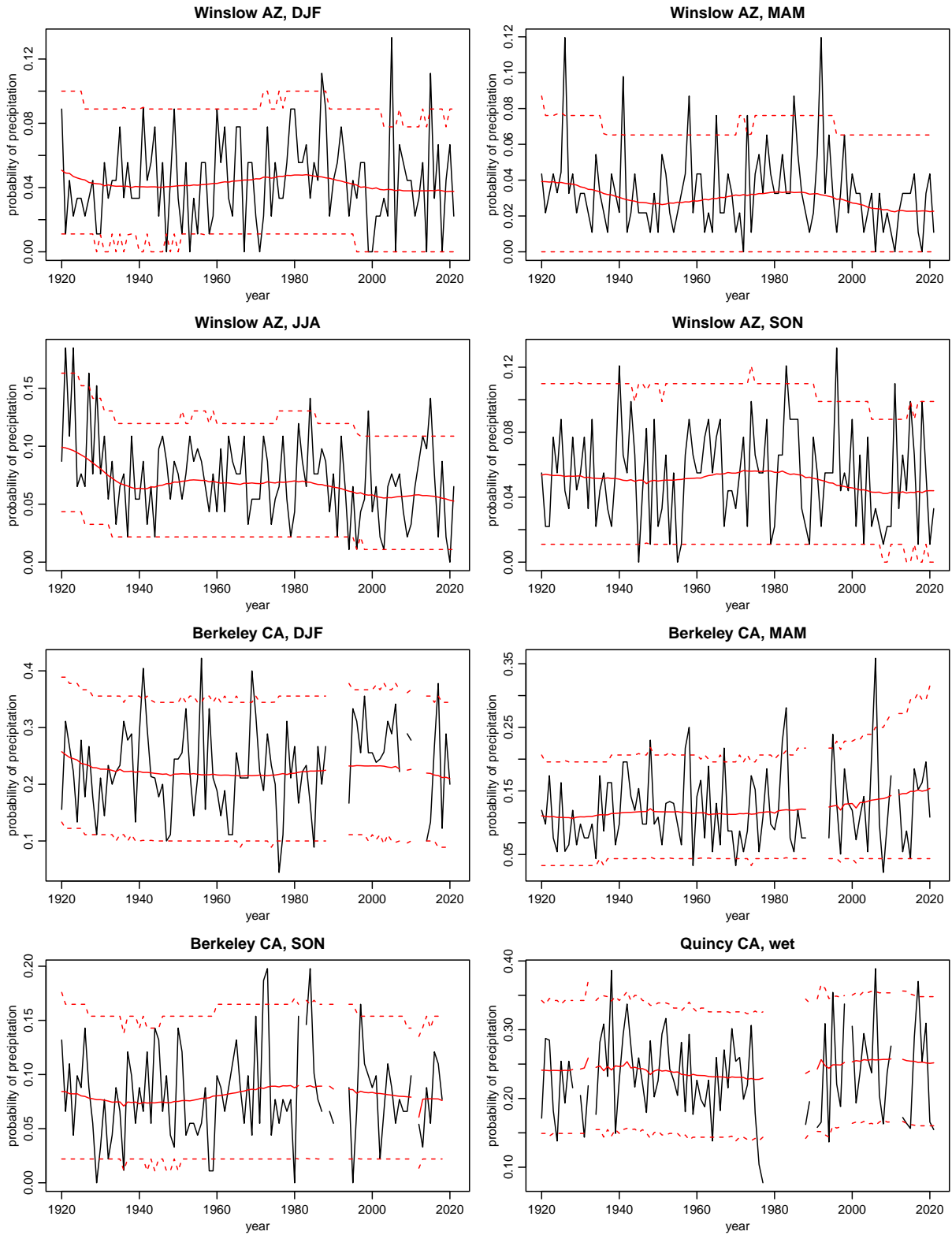}

\caption{\label{fig:diag_rain_year}Observed versus simulated probability of
non-negligible precipitation (greater than three mm) by year for the
eight location-season pairs, omitting years in which more than 25\%
of observations were missing. Simulated values are the mean over the
simulations, with 90\% predictive uncertainty bands.}
\end{figure}

\begin{figure}
\includegraphics[scale=0.75]{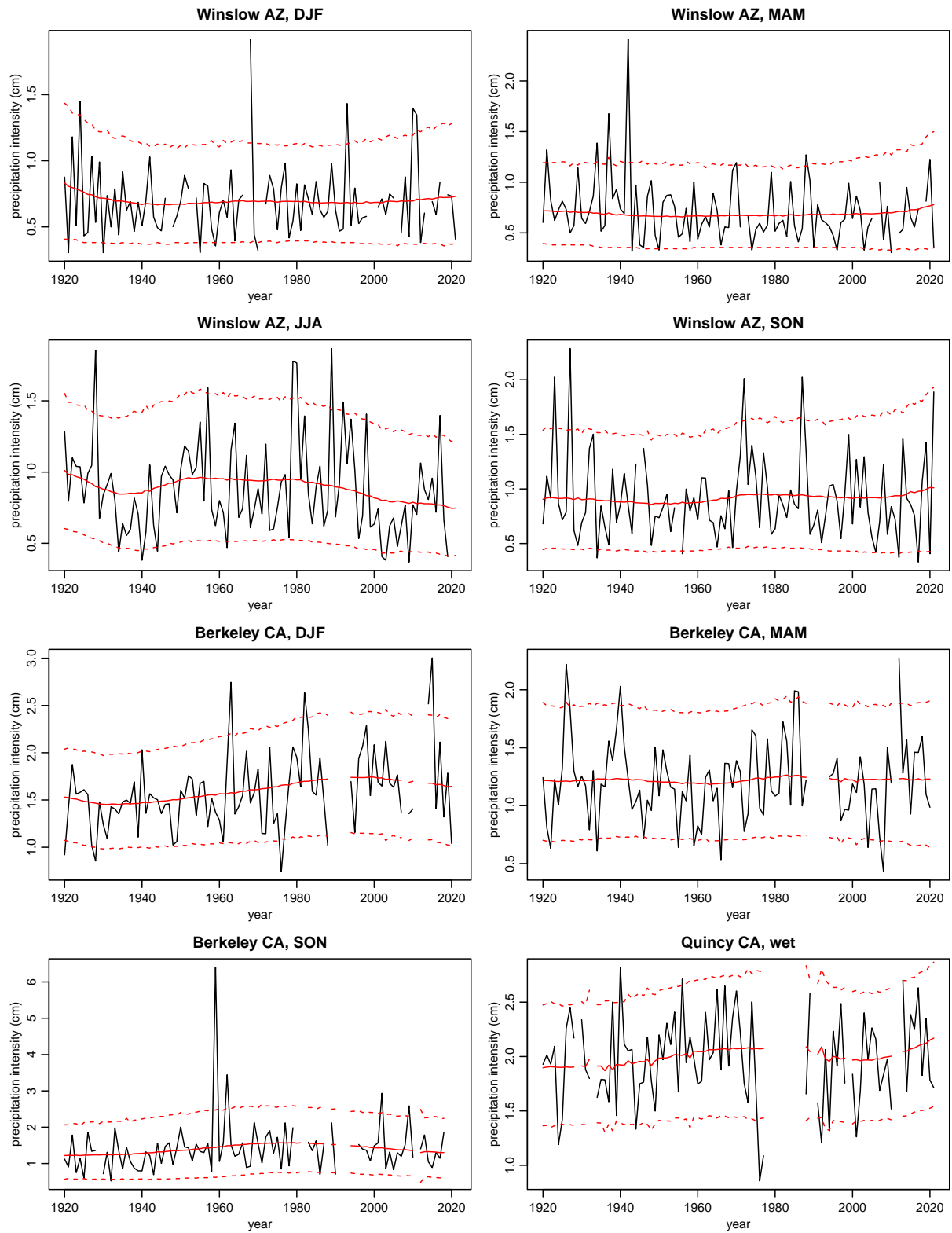}

\caption{\label{fig:diag_int_year}Observed versus simulated precipitation
intensity (mean of precipitation over days with non-negligible precipitation
(greater than three mm)) by year for the eight location-season pairs,
omitting years in which more than 25\% of observations were missing
(and omitting observed values for years with no days with non-negligible
precipitation). Simulated values are the mean over the simulations,
with 90\% predictive uncertainty bands. }
\end{figure}

\begin{figure}
\includegraphics[scale=0.75]{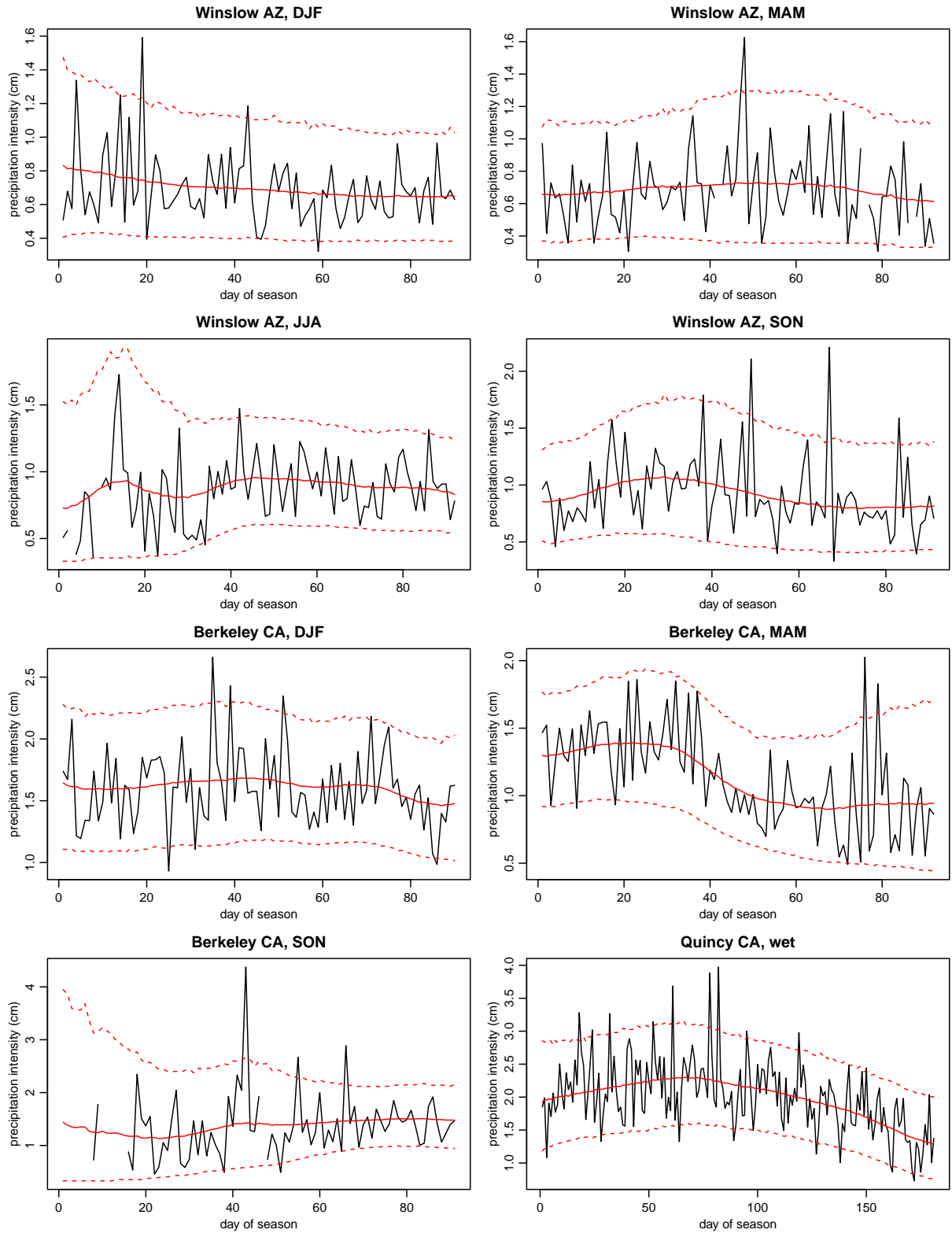}

\caption{\label{fig:diag_int_seas}Observed versus simulated precipitation
intensity (mean of precipitation over days with non-negligible precipitation
(greater than three mm)) within season (by day of season) for the
eight location-season pairs, omitting days of the season in which
more than 25\% of observations were missing (and omitting observed
values for days of the year with no days with non-negligible precipitation).
Simulated values are the mean over the simulations, with 90\% predictive
uncertainty bands. }
\end{figure}

\begin{figure}
\includegraphics[scale=0.75]{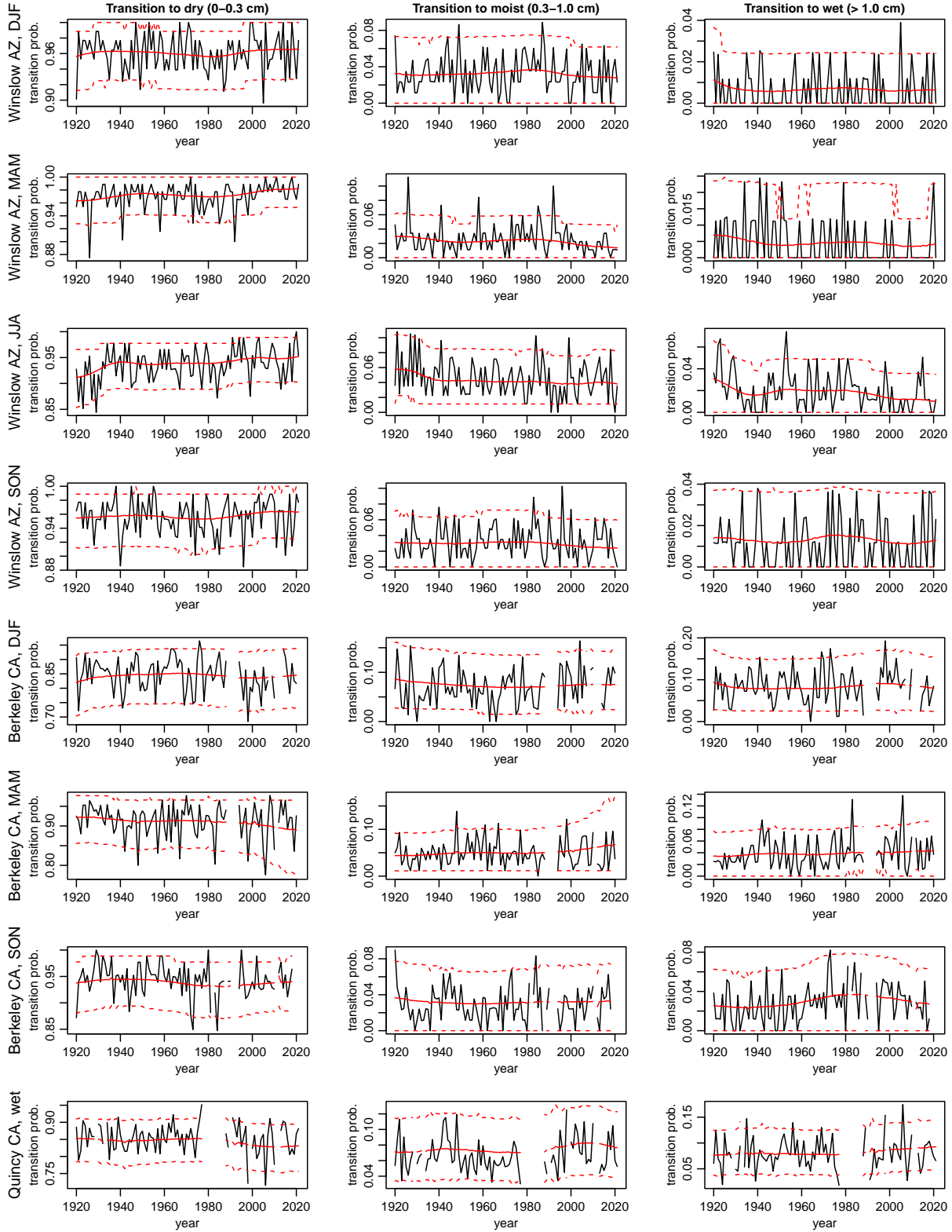}

\caption{\label{fig:diag_trans_from_dry_year}Observed versus simulated probability
of dry (0-0.3 cm; first column), moist (0.3-1 cm; second column),
or wet (>1 cm; third column) day given the previous day was dry (0-0.3
cm) by year, for the eight location-season pairs (rows). Years in
which more than 25\% of observations were missing are omitted. Simulated
values are the mean over the simulations, with 90\% predictive uncertainty
bands. }
\end{figure}

\begin{figure}
\includegraphics[scale=0.75]{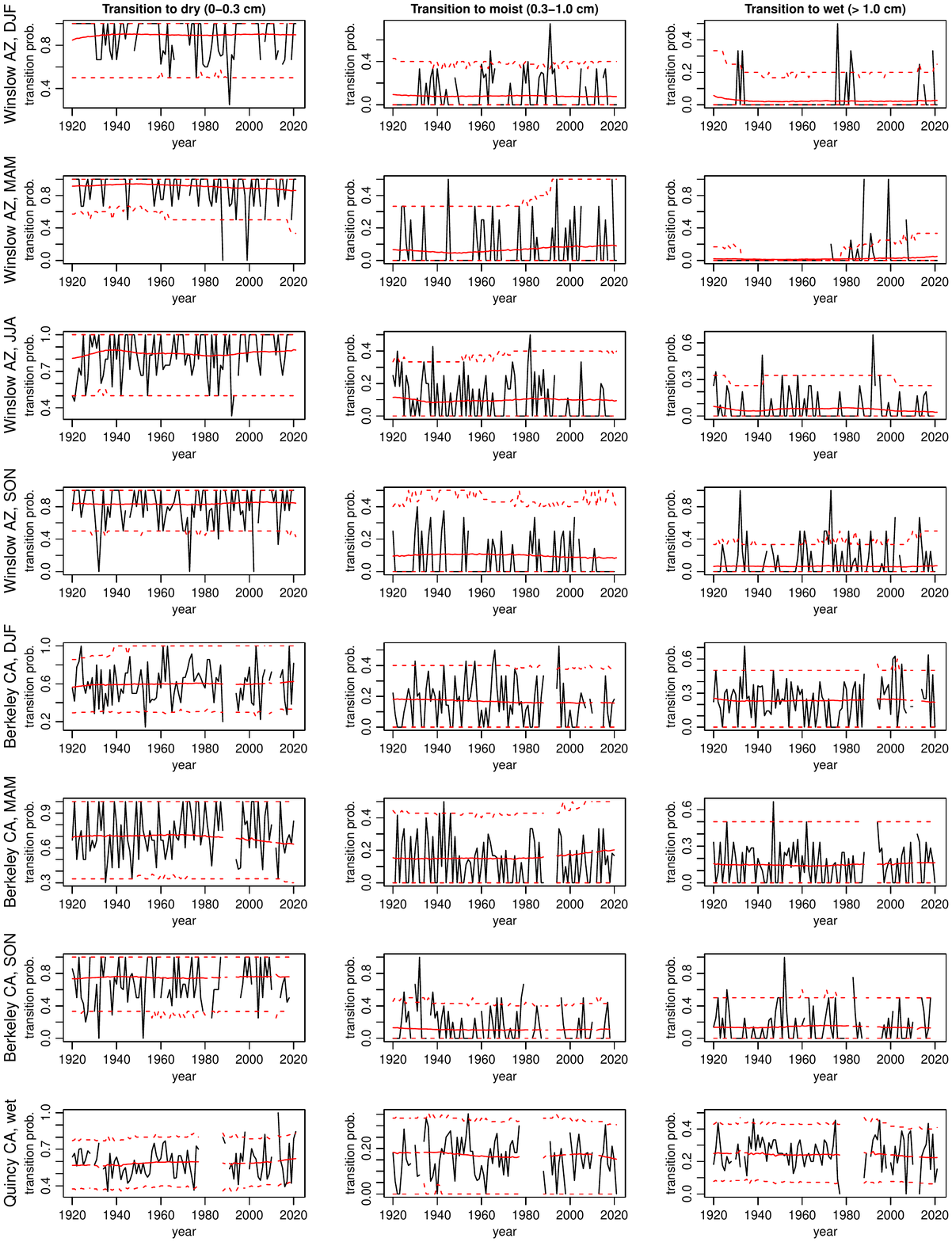}

\caption{\label{fig:diag_trans_from_moist_year}Observed versus simulated probability
of dry (0-0.3 cm; first column), moist (0.3-1 cm; second column),
or wet (>1 cm; third column) day given the previous day was moist
(0.3-1 cm) by year, for the eight location-season pairs (rows). Years
in which more than 25\% of observations were missing are omitted (as
are values for years in which there are no moist days). Simulated
values are the mean over the simulations, with 90\% predictive uncertainty
bands. Missing values can occur when there are no moist days in a
year.}
\end{figure}

\begin{figure}
\includegraphics[scale=0.75]{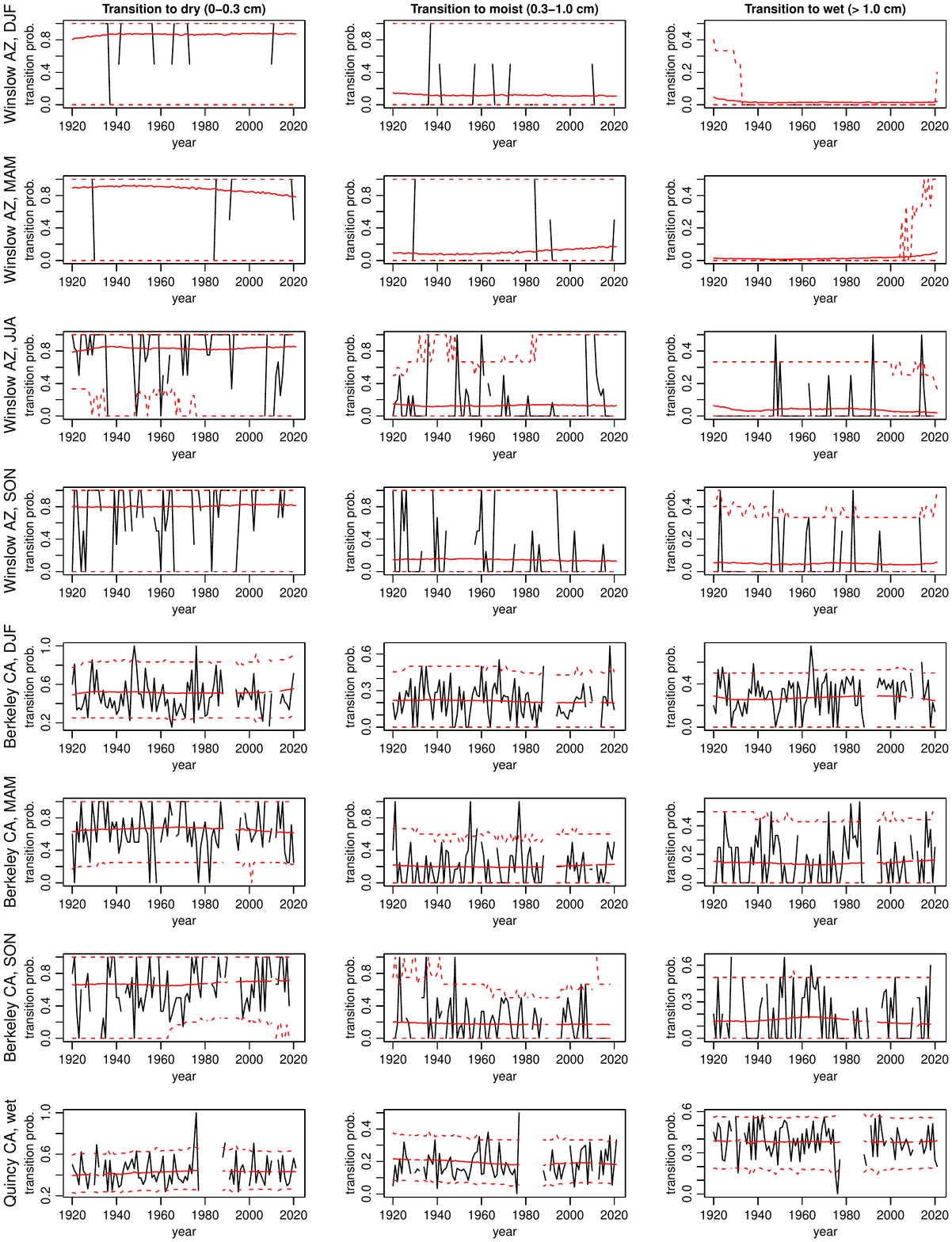}

\caption{\label{fig:diag_trans_from_wet_year}Observed versus simulated probability
of dry (0-0.3 cm; first column), moist (0.3-1 cm; second column),
or wet (>1 cm; third column) day given the previous day was wet (>1
cm) by year, for the eight location-season pairs (rows). Years in
which more than 25\% of observations were missing are omitted (as
are values for years in which there are no wet days). Simulated values
are the mean over the simulations, with 90\% predictive uncertainty
bands. }
\end{figure}

\begin{figure}
\includegraphics[scale=0.75]{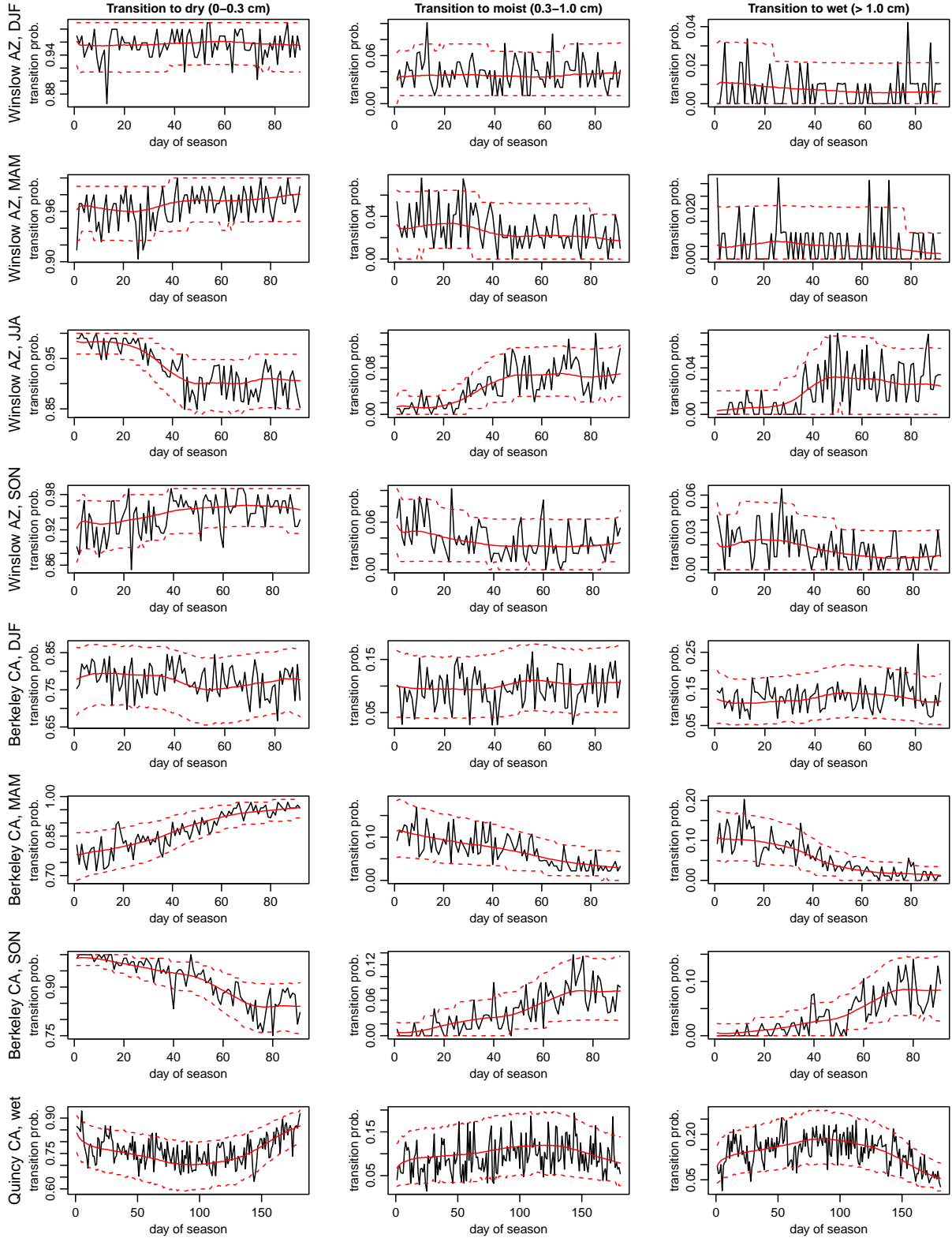}

\caption{\label{fig:diag_trans_from_dry_seas}Observed versus simulated probability
of dry (0-0.3 cm; first column), moist (0.3-1 cm; second column),
or wet (>1 cm; third column) day given the previous day was dry (0-0.3
cm) within season (by day of season), for the eight location-season
pairs (rows). Simulated values are the mean over the simulations,
with 90\% predictive uncertainty bands.}
\end{figure}

\begin{figure}
\includegraphics[scale=0.75]{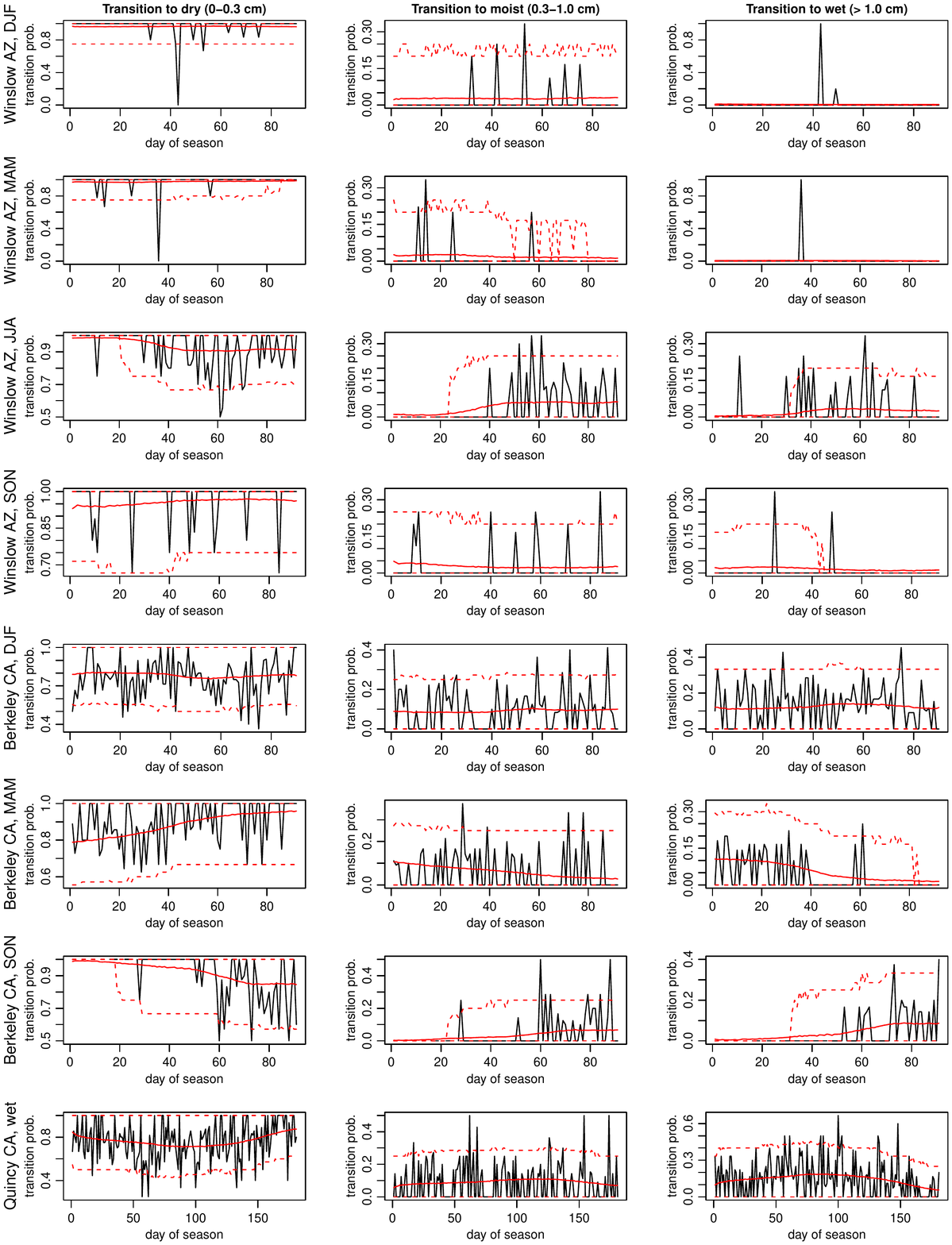}

\caption{\label{fig:diag_trans_from_moist_seas}Observed versus simulated probability
of dry (0-0.3 cm; first column), moist (0.3-1 cm; second column),
or wet (>1 cm; third column) day given the previous day was moist
(0.3-1 cm) within season (by day of season), for the eight location-season
pairs (rows). Days of season in which more than 25\% of observations
were missing are omitted (as are values for days of season in which
there are no moist days). Simulated values are the mean over the simulations,
with 90\% predictive uncertainty bands. }
\end{figure}

\begin{figure}
\includegraphics[scale=0.75]{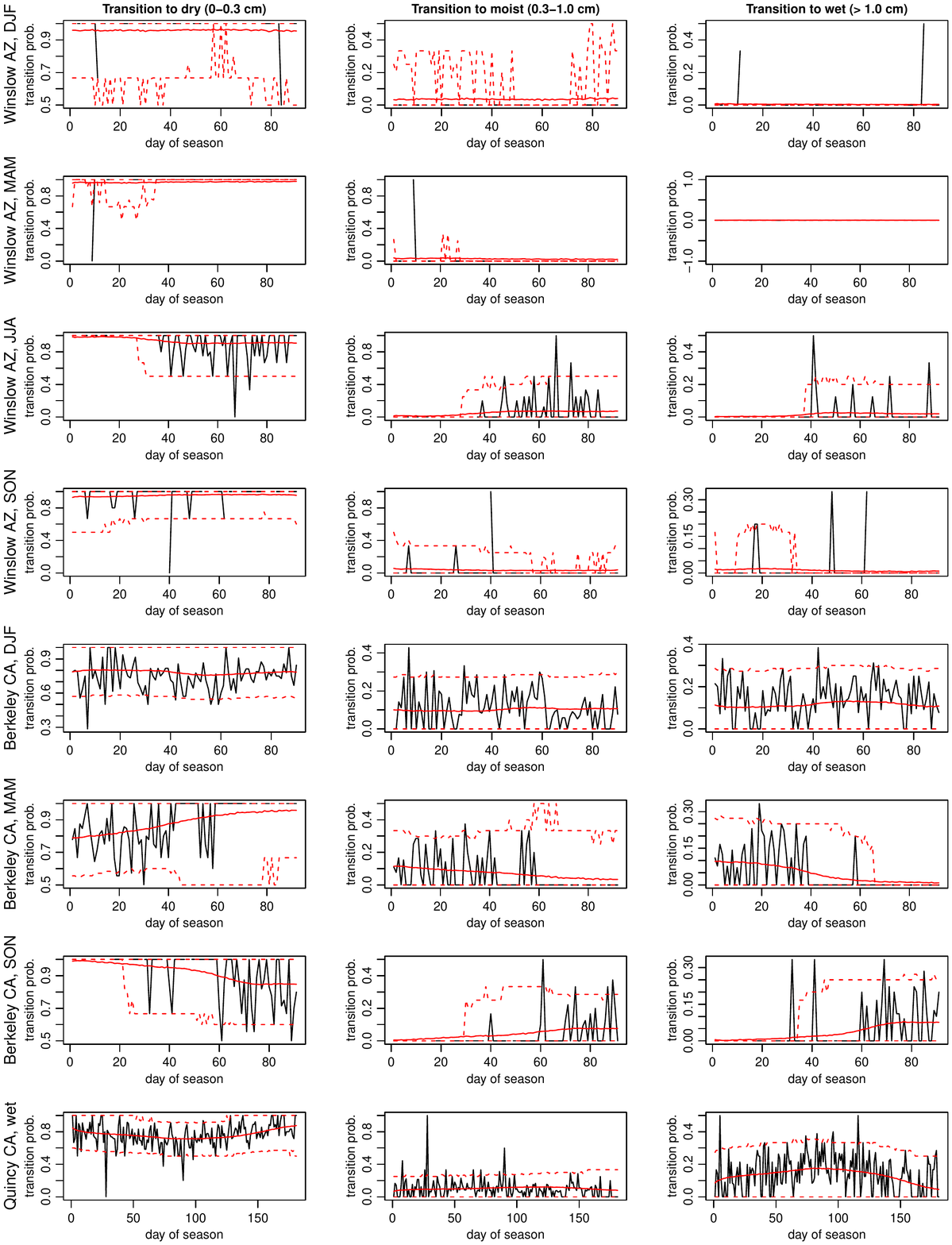}

\caption{\label{fig:diag_trans_from_wet_seas}Observed versus simulated probability
of dry (0-0.3 cm; first column), moist (0.3-1 cm; second column),
or wet (>1 cm; third column) day given the previous day was wet (>1
cm) within season (by day of season), for the eight location-season
pairs (rows). Days of season in which more than 25\% of observations
were missing are omitted (as are values for days of season in which
there are no wet days). Simulated values are the mean over the simulations,
with 90\% predictive uncertainty bands. }
\end{figure}

\begin{figure}
\includegraphics[scale=0.75]{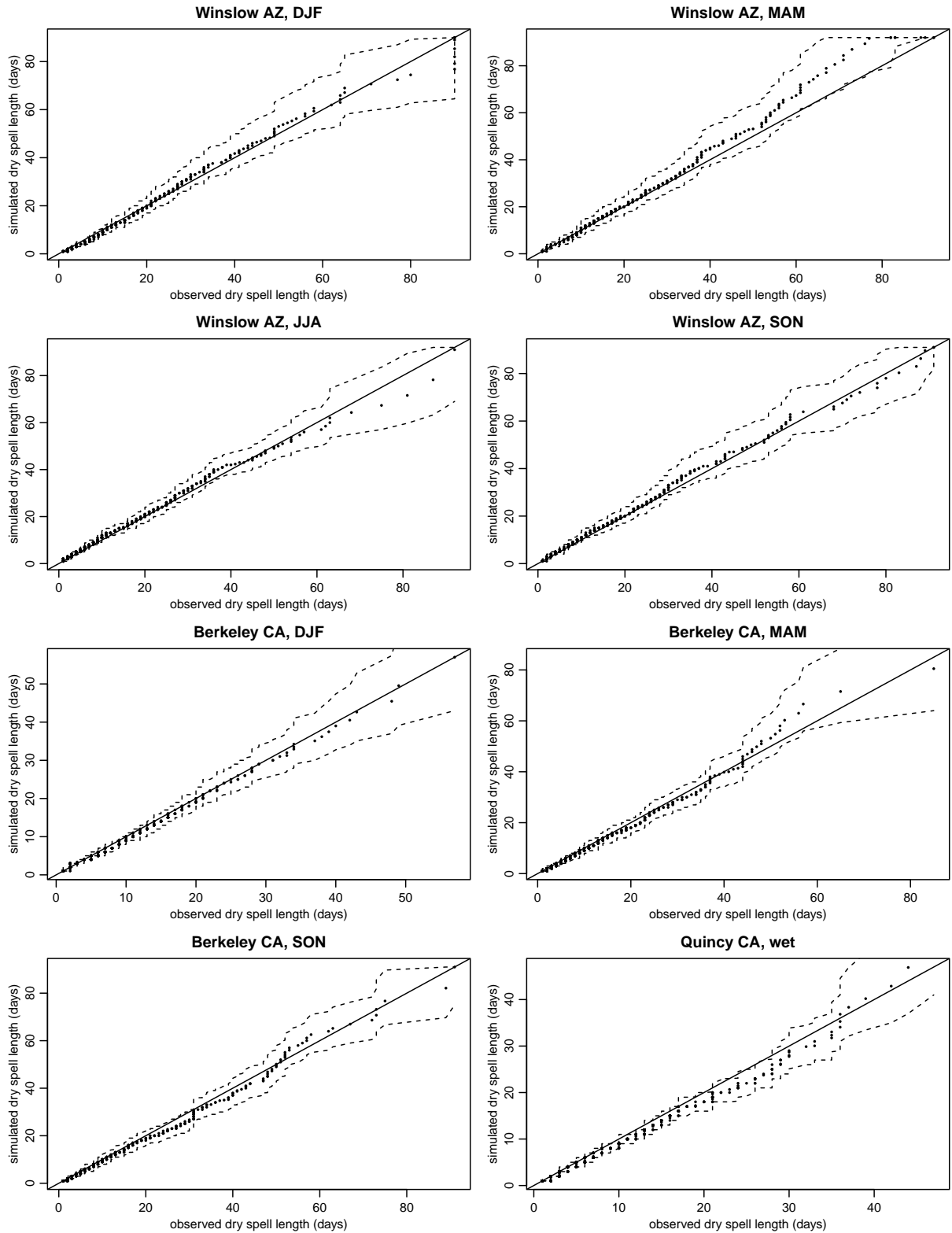}

\caption{\label{fig:diag_dsl_qq}Median and 95\% predictive intervals for Q-Q
comparisons of the simulated versus observed dry spell lengths for
the eight location-season pairs (rows). The median and 95\% intervals
are taken over Q-Q comparisons of each simulated time series against
the observed values. If simulated values are consistent with the observed
values, we expect the points (the median values) near the 1:1 line
and the intervals to generally cover the 1:1 line.}
\end{figure}

Figs. \ref{fig:diag_total}-\ref{fig:diag_dsl_qq} (Appendix) display
various additional diagnostic plots, showing the observed values compared
to the simulated values for total seasonal precipitation, probability
of precipitation, precipitation intensity, probabilities of discretized
precipitation given the previous day's discretized precipitation,
and Q-Q plots for dry spell length. In general, the simulated values
match the observed values well when accounting for stochasticity in
the observed values. The simulation uncertainty intervals do a reasonable
job of covering the observations, with some indications of overcoverage,
suggesting that the simulated values are a bit too variable. This
is perhaps not surprising as the simulations are probabilistic, while
the real data are generated by a physical system with embedded constraints. 

\subsection{Case studies\label{subsec:Case-Studies}}

\begin{figure}
\includegraphics[scale=0.75]{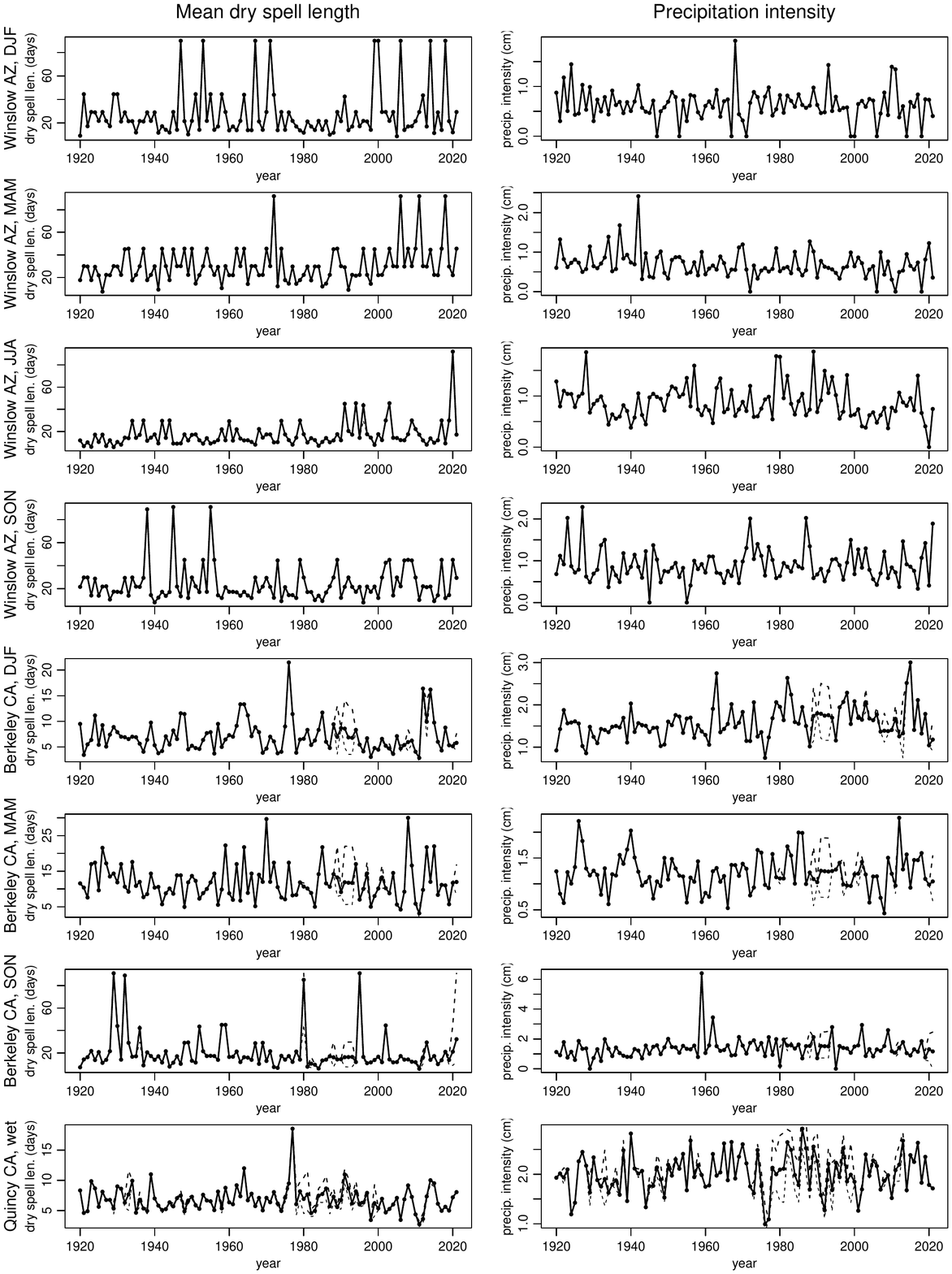}

\caption{\label{fig:Mean-dry-spell}Mean dry spell length (left column) and
precipitation intensity (right column) by year for the eight location-season
pairs. For years with any missing values, the dry spell length and
intensity are the mean over the posterior imputations, with 5\% and
95\% bounds shown in dashed lines. }
\end{figure}

Fig. \ref{fig:Mean-dry-spell} shows the mean dry spell length and
precipitation intensity by year for each of the eight location-season
pairs, with imputation uncertainty. The plots suggest a low signal-to-noise
ratio.

\subsubsection{Arizona drought}

\begin{figure}
\includegraphics[scale=0.75]{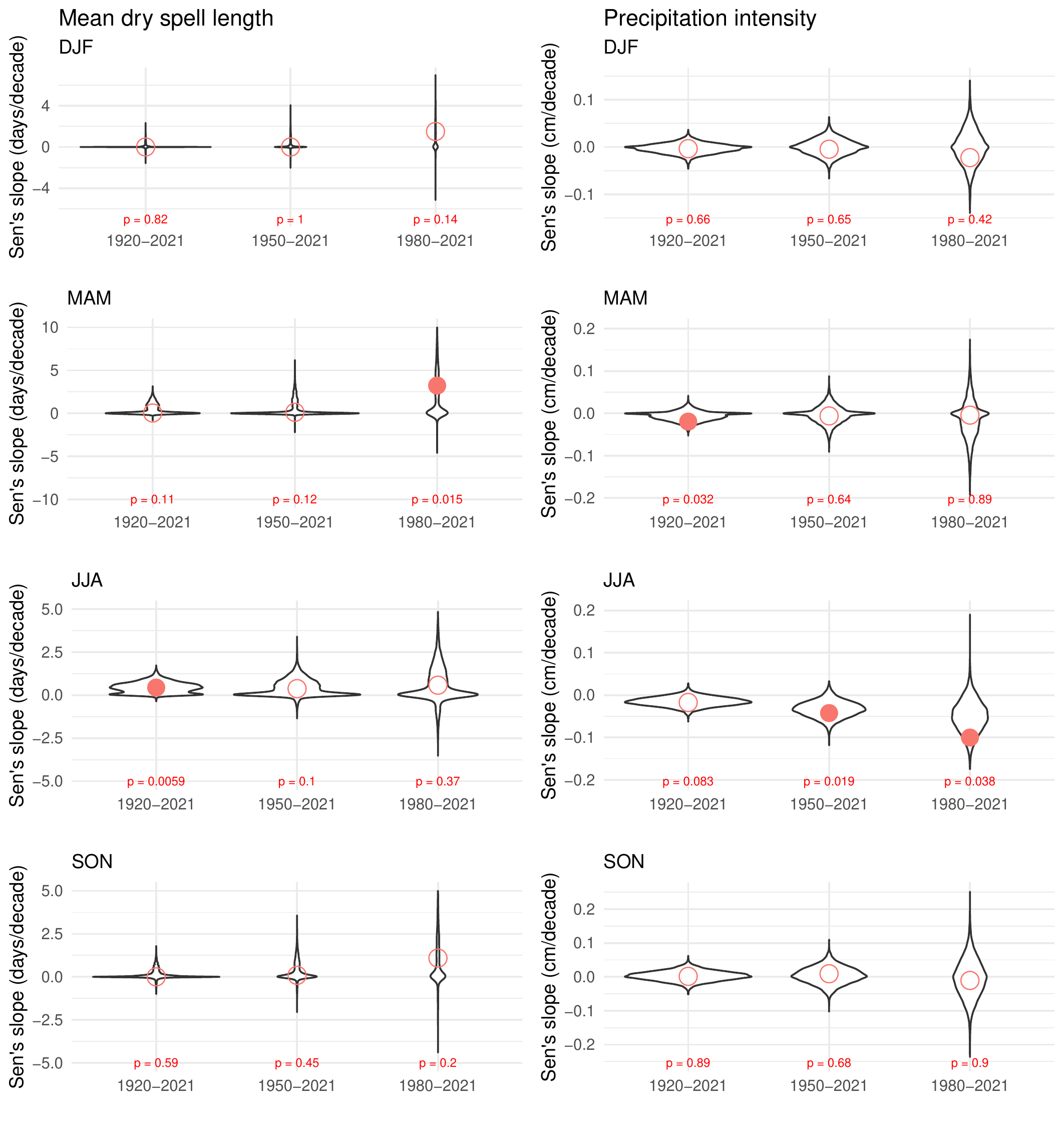}

\caption{\label{fig:Sen's-slope-estimates}Sen's slope estimates (scaled to
be on a per-decade basis) for four seasons (rows: DJF, MAM, JJA, SON)
for Winslow based on imputation (red circles; filled circles for estimates
with p-values less than 0.05) and the Bayesian posterior (violin plots)
for the mean dry spell length (left) and precipitation intensity (right),
with p-values for the imputation-based frequentist analysis reported
in text in the figure. For the dry spell length (left), the violin
plots reflect spikes at zero caused by the spell length being discrete
and Sen's slope being the median of pairwise scaled differences between
the spell length in different years, with differences of zero being
common.}
\end{figure}

Fig. \ref{fig:Sen's-slope-estimates} presents trend results for Winslow
for DJF, MAM, JJA and SON. We see some evidence for trends in both
dry spell length (increasing) and precipitation intensity (decreasing)
for MAM and JJA for some of the time periods. One cautionary note
is that MCMC mixing for JJA for the longer time periods is problematic
(Table \ref{tab:MCMC-diagnostics-(and}), although we don't expect
this to have much impact on the imputation-based analysis given the
negligible missing data for this time series.

\subsubsection{California winter dry spells}

\begin{figure}
\includegraphics[scale=0.75]{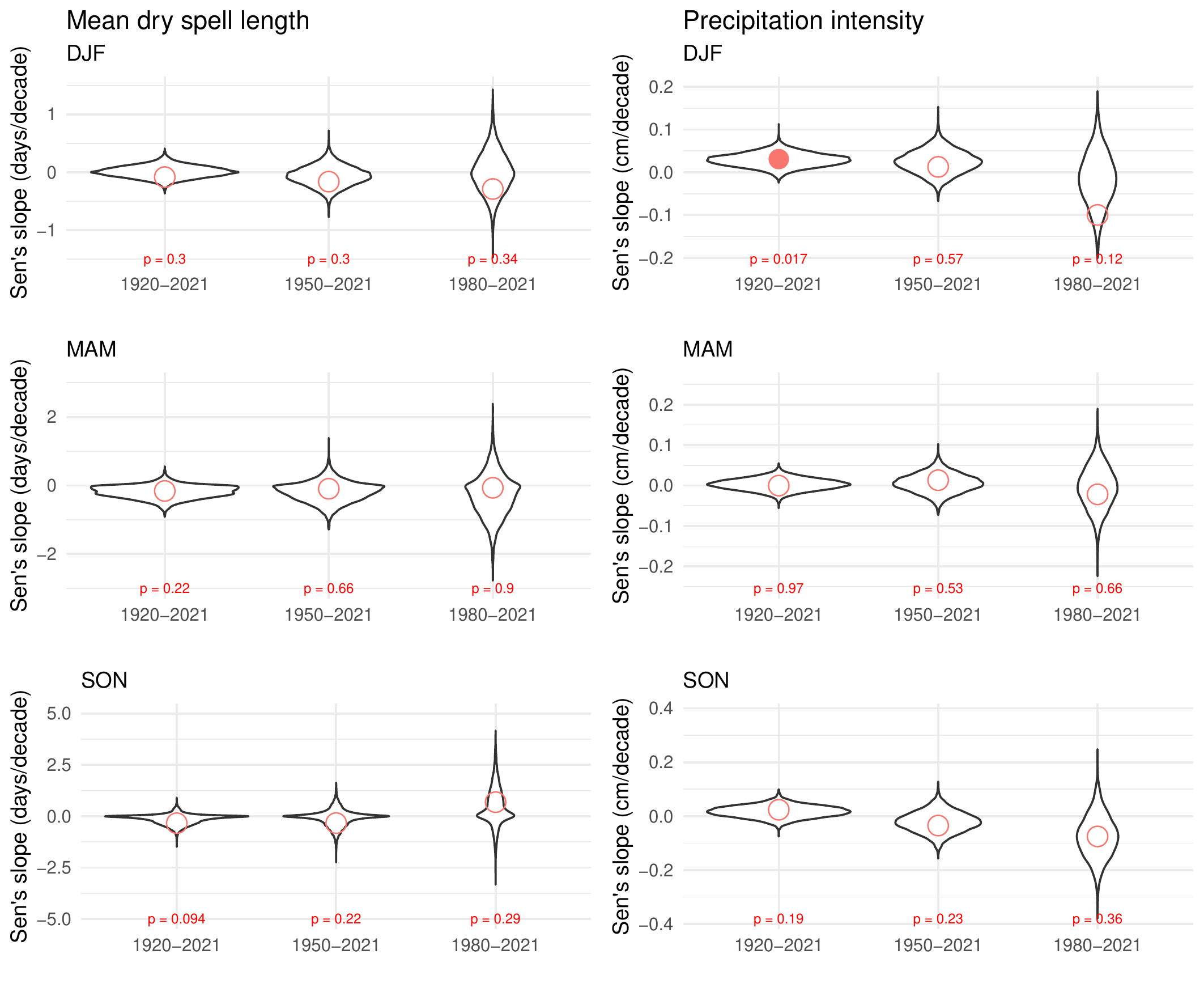}

\caption{\label{fig:Sen's-slope-estimates-1}Sen's slope estimates (scaled
to be on a per-decade basis) for three seasons (rows: DJF, MAM, SON)
for Berkeley based on imputation (red circles; filled circles for
estimates with p-values less than 0.05) and the Bayesian posterior
(violin plots) for the mean dry spell length (left) and precipitation
intensity (right), with p-values for the imputation-based frequentist
analysis reported in text in the figure. For the dry spell length
(left), the violin plots reflect spikes at zero caused by the spell
length being discrete and Sen's slope being the median of pairwise
scaled differences between the spell length in different years, with
differences of zero being common.}
\end{figure}

\begin{figure}
\includegraphics[scale=0.75]{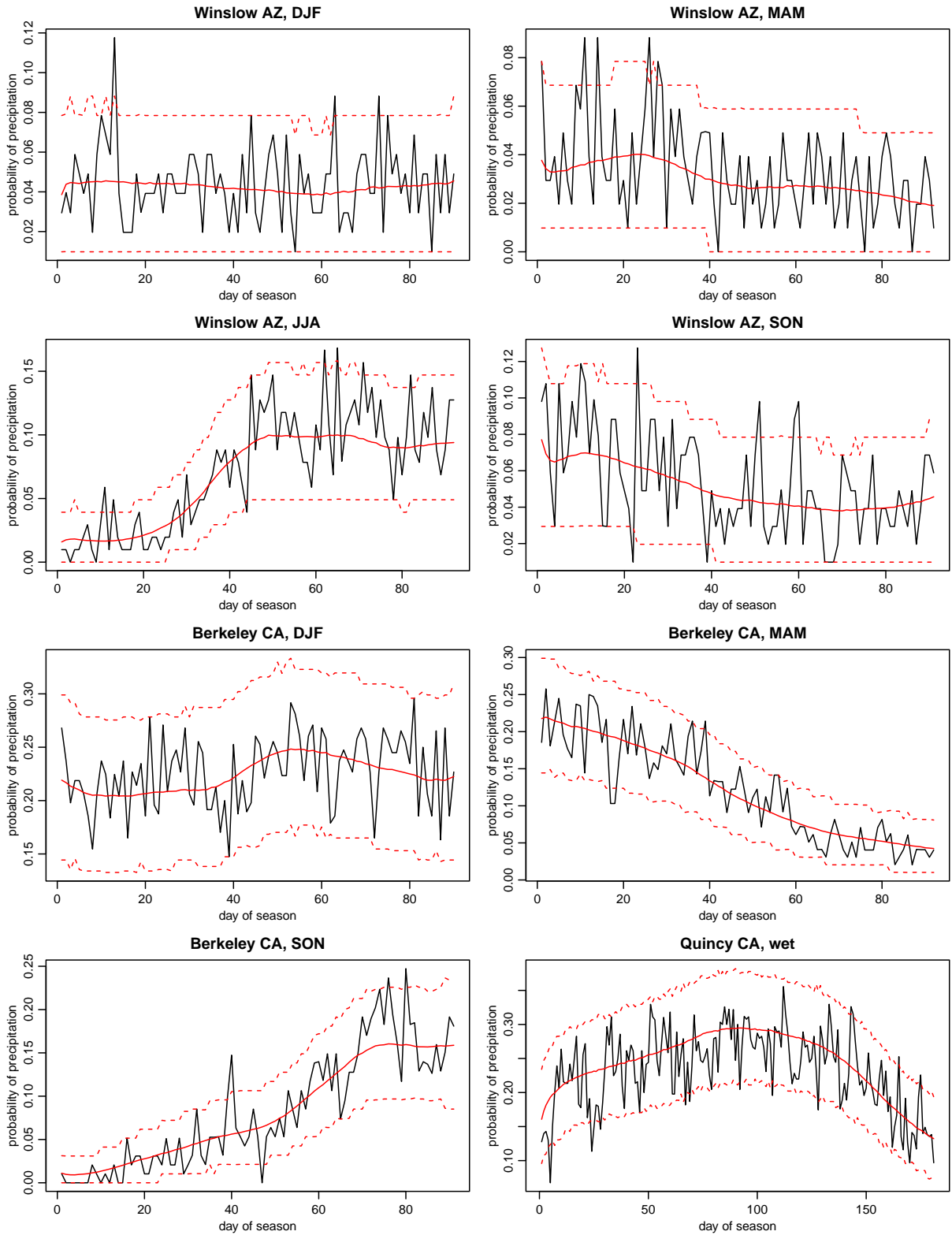}

\caption{\label{fig:diag_rain_seas}Observed versus simulated probability of
non-negligible precipitation (greater than three mm) within season
(by day of season) for the eight location-season pairs, omitting days
of the year in which more than 25\% of observations were missing.
Simulated values are the mean over the simulations, with 90\% predictive
uncertainty bands.}
\end{figure}

Fig. \ref{fig:Sen's-slope-estimates-1} presents trend results for
Berkeley for DJF, MAM, and SON. We omit JJA because of the extreme
dry season during those months, with very little precipitation (e.g.,
see Fig. \ref{fig:diag_rain_seas} where the probability of precipitation
at the end of MAM and beginning of SON is very low). There is little
evidence for trend over any of the three time periods for either metric,
apart from an increase in precipitation intensity for DJF over 1920-2021.
This suggests that, at least as measured by the dry spell length and
precipitation intensity metrics, we do not see evidence in support
of trend in blocking behavior or increased within-season variability
of precipitation.

\subsubsection{California mountain precipitation}

\begin{figure}
\includegraphics[scale=0.6]{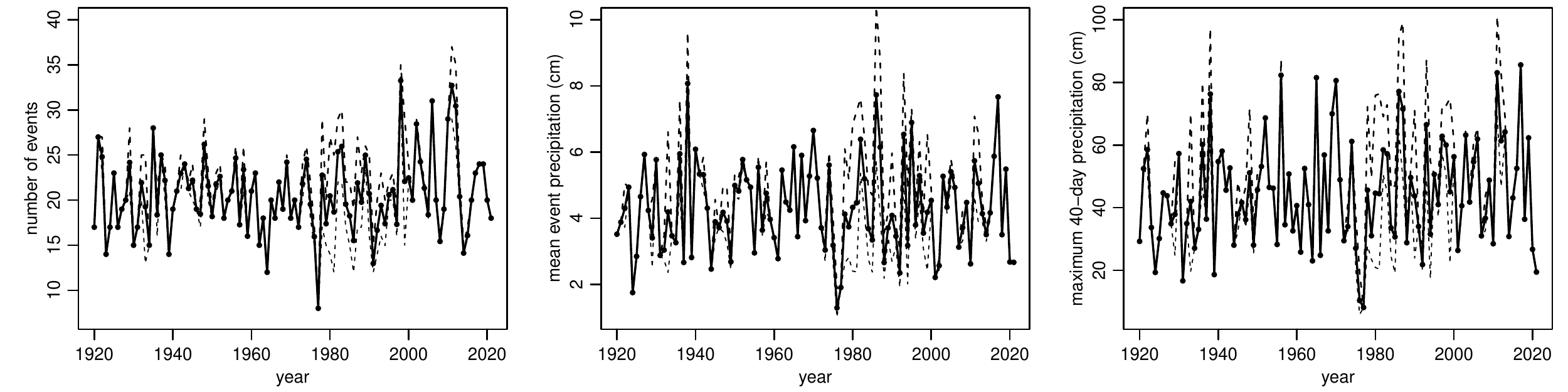}

\caption{\label{fig:Number-of-wet}Number of wet spells (left), mean wet spell
precipitation (cm) (middle) and maximum 40-day precipitation (cm)
(right) by year for Quincy, California. For years with any missing
values, the point values are the mean over the posterior imputations,
with 5\% and 95\% bounds shown in dashed lines. }
\end{figure}

Fig. \ref{fig:Number-of-wet} shows the number of wet spells, mean
spell-level precipitation, and maximum 40-day precipitation for Quincy,
with imputation uncertainty. The plots suggest a low signal-to-noise
ratio.

\begin{figure}
\includegraphics[scale=0.75]{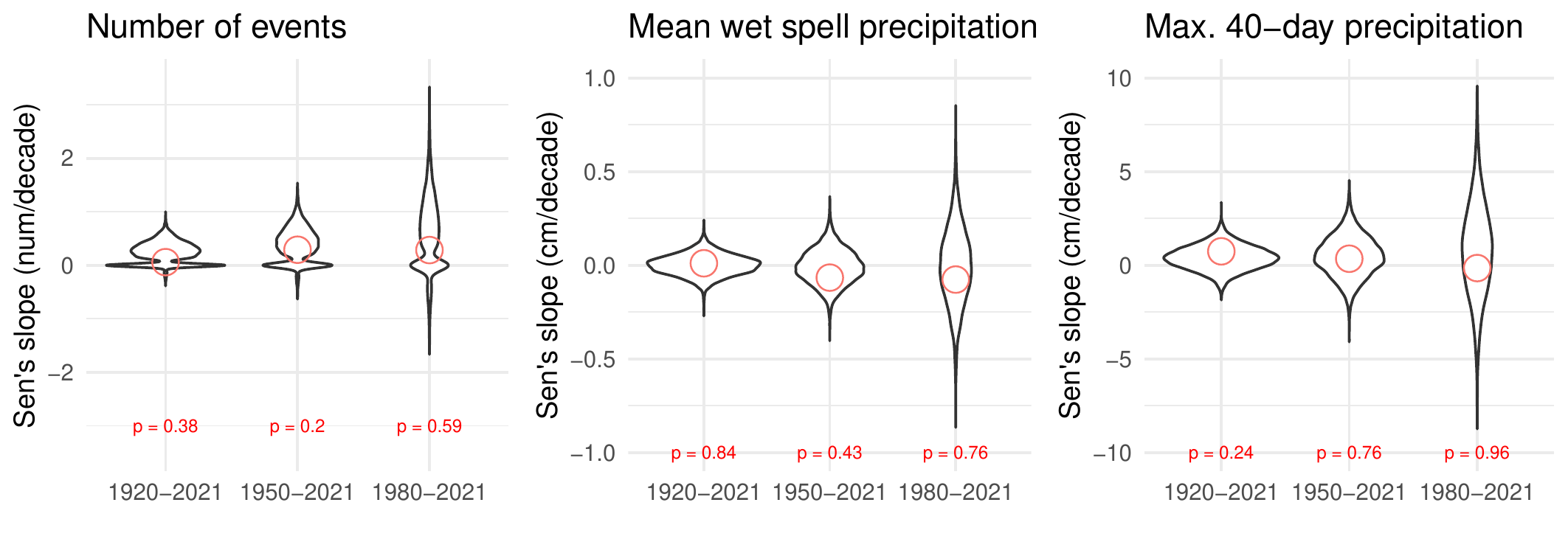}

\caption{\label{fig:Sen's-slope-estimates-2}Sen's slope estimates (scaled
to be on a per-decade basis) based on imputation (circles) and the
Bayesian posterior (violin plots) for the number of precipitation
events (left), mean precipitation per event (center) and maximum 40-day
precipitation in a season (right), with p-values for the imputation-based
frequentist analysis reported in text in the figure. For the number
of events (left), the violin plots reflect spikes at zero caused by
the number of events being discrete and Sen's slope being the median
of pairwise scaled differences between the number of events in different
years, with differences of zero being common.}
\end{figure}

Fig. \ref{fig:Sen's-slope-estimates-2} presents trend results for
Quincy. There is little evidence for trend over any of the three time
periods or three precipitation metrics. Thus we do not see evidence
that the precipitation pattern resulting in the Oroville Dam event
was related to a long-term trend in the pattern of precipitation events
as measured by our metrics. Of course our metrics do not capture other
potentially-relevant aspects of climate relevant to the event, such
as rain-on-snow events that can cause a burst of snowmelt. 

\section{Discussion}

We have presented a stochastic weather generator-based approach to
inference for trends in daily precipitation patterns, focusing not
on traditional short-term extremes but on multi-day and longer precipitation
patterns that may have important impacts or reflect changing climate.
The approach focuses on careful modeling of precipitation time series
using hidden Markov models in a Bayesian framework. As such, it is
suitable for analyzing single time series, which has strengths and
limitations. By focusing on a single time series, we have more confidence
in the ability of the model to characterize the underlying precipitation
dynamics than a more complicated spatio-temporal model, and the method
can be used in locations without a network of stations. But it leaves
aside the possibility of bringing in additional information that might
increase the signal-to-noise ratio, including external information
such as radar or satellite data or even other weather variables likely
available from the location of interest.

We have used the approach to make inference in two ways. The first
is to use the model as an imputation model and then use classical
statistical methods to do inference, adjusting for the imputation
uncertainty. The second is to use the model within a fully Bayesian
framework, doing infererence based on the posterior predictive distribution.
The latter of course relies more heavily on the statistical model
sufficiently capturing the feature of the time series, but it has
the benefit of fully Bayesian joint inference for any quantities of
interest. The former likely has limited dependence on the model, particularly
when missing data are infrequent, while still avoiding the difficulties
that missing data pose, both in terms of potential bias from missingness
(in particular missing at random mechanisms with respect to year and
day of year) and in terms of quantities that cannot be reliably calculated
in the face of missing data (such as dry spell lengths). 

In our case studies, we find mixed evidence for trends in precipitation
patterns based on individual stations, with most of the location-season
pairs showing little evidence of trend. This is perhaps not surprising
given the high entropy of precipitation. In future work, we will apply
the model to large sets of station data (see \citealp{zhang2021five}
for such an approach). Unlike other such analyses, we will not be
as limited by missing data and by the need to determine criteria for
removing years of data or making assumptions about the missing values.
However, this future work will still suffer from lack of a coherent
joint approach to uncertainty in a spatial context. This is a long-standing
shortcoming of spatio-temporal analyses. While there is statistical
literature on spatial multiple testing \citep[e.g.,][]{sun2015false,risser2019spatially},
there is not a well-developed general methodology for doing so with
spatially-correlated p-values, and Bayesian approaches rely heavily
on the statistical model being fit for purpose, a particularly strong
assumption in real world spatial settings with complicated spatial
nonstationarity. Furthermore, for the questions of interest, it's
not clear how much additional information is provided by nearby locations
given the spatial extent of weather phenomena. For example, it's not
clear that an analysis of a region of California encompassing Berkeley
would give substantially more informative results about potential
effects of blocking and precipitation variability given that the same
weather events affect large spatial areas (particularly where large-scale
frontal systems are concerned). 

One potential approach that retains the benefit of some simplicity
but allows use of multiple locations would be a joint imputation model
that seeks to impute missing data for all stations within a spatial
region. (Using the methodology presented here, doing imputation individually
at multiple stations is not appropriate as the imputed values at a
given time will be imputed independently and not reflect the true
spatial structure.) Of course such a joint imputation model will rely
on the quality of the underlying statistical model with respect to
the spatial structure of precipitation. Again, however, by focusing
on imputation, one can hope to limit the influence of the model, while
still allowing for borrowing of strength spatially without as much
concern that the patterns of missingness in space and time may bias
results.

\subsection*{Acknowledgments}

This research was supported by the Director, Office of Science, Office
of Biological and Environmental Research of the U.S. Department of
Energy under Contract No. DE-AC02-05CH11231. I thank my CASCADE colleagues
for suggestions, in particular Travis O'Brien, Mark Risser, and Michael
Wehner.

This document was prepared as an account of work sponsored by the
U.S. government. While this document is believed to contain correct
information, neither the U.S. government nor any agency thereof, nor
the Regents of the University of California, nor any of their employees,
makes any warranty, express or implied, or assumes any legal re- sponsibility
for the accuracy, completeness, or usefulness of any information,
apparatus, product, or process disclosed, or represents that its use
would not infringe privately owned rights. Reference herein to any
specific commercial product, process, or service by its trade name,
trademark, manufacturer, or otherwise, does not necessarily constitute
or imply its endorsement, recommendation, or favoring by the U.S.
government or any agency thereof, or the Regents of the University
of California. The views and opinions of authors expressed herein
do not necessarily state or reflect those of the U.S. government or
any agency thereof or the Regents of the University of California.

\bibliographystyle{chicago}
\bibliography{refs}

\part*{\protect\pagebreak Supplemental figures and tables}

\begin{table}[H]
\caption{\label{tab:Median-(99th-percentile)}Median (99th percentile)$\hat{R}$
MCMC diagnostic across 102 yearly values (or \textasciitilde 90 day
of season values) for various quantities for the eight location-seasons.}

\begin{tabular}{|>{\centering}p{1.75cm}|>{\centering}p{1cm}|>{\centering}p{1cm}|>{\centering}p{1cm}|>{\centering}p{1cm}|>{\centering}p{1cm}|>{\centering}p{1cm}|>{\centering}p{1cm}|>{\centering}p{1cm}|>{\centering}p{1cm}|>{\centering}p{1cm}|}
\hline 
 &
{\scriptsize{}yearly prob. rain > 3 mm} &
{\scriptsize{}yearly prob. rain > 10 mm} &
{\scriptsize{}yearly prob. rain > 20 mm} &
{\scriptsize{}day of year prob. rain > 3 mm} &
{\scriptsize{}day of year prob. rain > 10 mm} &
{\scriptsize{}day of year prob. rain > 20 mm} &
{\scriptsize{}yearly prob. dry to dry} &
{\scriptsize{}yearly prob. wet to dry} &
{\scriptsize{}yearly mean dry spell length} &
{\scriptsize{}yearly precipitation intensity }\tabularnewline
\hline 
\hline 
{\scriptsize{}Winslow AZ, DJF} &
{\scriptsize{}1.001 (1.004)} &
{\scriptsize{}1.001 (1.002)} &
{\scriptsize{}1.000 (1.002)} &
{\scriptsize{}1.001 (1.004)} &
{\scriptsize{}1.000 (1.002)} &
{\scriptsize{}1.000 (1.002)} &
{\scriptsize{}1.001 (1.004)} &
{\scriptsize{}1.000 (1.003)} &
{\scriptsize{}1.001 (1.004)} &
{\scriptsize{}1.000 (1.002)}\tabularnewline
\hline 
{\scriptsize{}Winslow AZ, MAM} &
{\scriptsize{}1.000 (1.001)} &
{\scriptsize{}1.000 (1.002)} &
{\scriptsize{}1.000 (1.001)} &
{\scriptsize{}1.000 (1.001)} &
{\scriptsize{}1.000 (1.002)} &
{\scriptsize{}1.000 (1.001)} &
{\scriptsize{}1.000 (1.001)} &
{\scriptsize{}1.000 (1.001)} &
{\scriptsize{}1.000 (1.001)} &
{\scriptsize{}1.000 (1.001)}\tabularnewline
\hline 
{\scriptsize{}Winslow AZ, JJA} &
{\scriptsize{}1.001 (1.014)} &
{\scriptsize{}1.001 (1.009)} &
{\scriptsize{}1.001 (1.010)} &
{\scriptsize{}1.002 (1.008)} &
{\scriptsize{}1.001 (1.022)} &
{\scriptsize{}1.001 (1.014)} &
{\scriptsize{}1.002 (1.011)} &
{\scriptsize{}1.001 (1.007)} &
{\scriptsize{}1.002 (1.011)} &
{\scriptsize{}1.001 (1.009)}\tabularnewline
\hline 
{\scriptsize{}Winslow AZ, SON} &
{\scriptsize{}1.001 (1.002)} &
{\scriptsize{}1.000 (1.001)} &
{\scriptsize{}1.000 (1.001)} &
{\scriptsize{}1.001 (1.004)} &
{\scriptsize{}1.000 (1.002)} &
{\scriptsize{}1.000 (1.001)} &
{\scriptsize{}1.000 (1.002)} &
{\scriptsize{}1.001 (1.002)} &
{\scriptsize{}1.000 (1.002)} &
{\scriptsize{}1.000 (1.001)}\tabularnewline
\hline 
{\scriptsize{}Berkeley CA, DJF} &
{\scriptsize{}1.001 (1.004)} &
{\scriptsize{}1.001 (1.011)} &
{\scriptsize{}1.001 (1.006)} &
{\scriptsize{}1.002 (1.012)} &
{\scriptsize{}1.001 (1.015)} &
{\scriptsize{}1.001 (1.011)} &
{\scriptsize{}1.001 (1.008)} &
{\scriptsize{}1.001 (1.014)} &
{\scriptsize{}1.001 (1.007)} &
{\scriptsize{}1.001 (1.003)}\tabularnewline
\hline 
{\scriptsize{}Berkeley CA, MAM} &
{\scriptsize{}1.001 (1.003)} &
{\scriptsize{}1.001 (1.004)} &
{\scriptsize{}1.001 (1.002)} &
{\scriptsize{}1.002 (1.006)} &
{\scriptsize{}1.001 (1.006)} &
{\scriptsize{}1.001 (1.004)} &
{\scriptsize{}1.001 (1.004)} &
{\scriptsize{}1.001 (1.002)} &
{\scriptsize{}1.001 (1.004)} &
{\scriptsize{}1.001 (1.002)}\tabularnewline
\hline 
{\scriptsize{}Berkeley CA, SON} &
{\scriptsize{}1.000 (1.001)} &
{\scriptsize{}1.000 (1.001)} &
{\scriptsize{}1.000 (1.002)} &
{\scriptsize{}1.001 (1.008)} &
{\scriptsize{}1.000 (1.005)} &
{\scriptsize{}1.000 (1.005)} &
{\scriptsize{}1.000 (1.001)} &
{\scriptsize{}1.000 (1.002)} &
{\scriptsize{}1.000 (1.002)} &
{\scriptsize{}1.000 (1.001)}\tabularnewline
\hline 
{\scriptsize{}Quincy CA, wet (Nov-Apr)} &
{\scriptsize{}1.000 (1.002)} &
{\scriptsize{}1.000 (1.002)} &
{\scriptsize{}1.000 (1.002)} &
{\scriptsize{}1.000 (1.003)} &
{\scriptsize{}1.000 (1.003)} &
{\scriptsize{}1.000 (1.003)} &
{\scriptsize{}1.000 (1.002)} &
{\scriptsize{}1.000 (1.002)} &
{\scriptsize{}1.000 (1.001)} &
{\scriptsize{}1.000 (1.002)}\tabularnewline
\hline 
\end{tabular}
\end{table}

\end{document}